\documentclass[journal=jacsat,manuscript=article]{achemso}

\usepackage[version=3]{mhchem} 
\usepackage{gensymb}
\usepackage{chemformula} 
\usepackage[T1]{fontenc} 
\usepackage{textgreek} 
\usepackage{color,soul}
\usepackage[font=bf]{caption} 
\usepackage{lipsum}
\usepackage{multirow}
\renewcommand\hl[1]{#1} 

\SectionNumbersOn 
\setkeys{acs}{keywords = false}
\setkeys{acs}{articletitle = true}

\author{Elias Sebti}
\affiliation{Materials Department, University of California, Santa Barbara, California 93106, United States}
\altaffiliation{Materials Research Laboratory, University of California, Santa Barbara, California 93106, United States}

\author{Hayden A. Evans}
\affiliation{Center for Neutron Research, National Institute of Standards and Technology, Gaithersburg, Maryland 20899, USA}

\author{Hengning Chen}
\affiliation{Department of Materials Science and Engineering, National University of
Singapore, 9 Engineering Drive 1, 117575, Singapore}

\author{Peter M. Richardson}
\affiliation{Materials Department, University of California, Santa Barbara, California 93106, United States}
\altaffiliation{Materials Research Laboratory, University of California, Santa Barbara, California 93106, United States}

\author{Kelly M. White}
\affiliation{Chemistry and Biochemistry Department, University of California, Santa Barbara, California 93106, United States}
\altaffiliation{Materials Research Laboratory, University of California, Santa Barbara, California 93106, United States}

\author{Raynald Giovine}
\affiliation{Materials Department, University of California, Santa Barbara, California 93106, United States}
\altaffiliation{Materials Research Laboratory, University of California, Santa Barbara, California 93106, United States}

\author{Krishna Prasad Koirala}
\affiliation{Physical and Computational Sciences Directorate, Pacific Northwest National Laboratory, Richland, WA, 99352, USA}

\author{Yaobin Xu}
\affiliation{Environmental Molecular Sciences Laboratory, Pacific Northwest National Laboratory, Richland, WA, 99352, USA}

\author{Eliovardo Gonzalez-Correa}
\affiliation{Materials Department, University of California, Santa Barbara, California 93106, United States}
\altaffiliation{Materials Research Laboratory, University of California, Santa Barbara, California 93106, United States}

\author{Chongmin Wang}
\affiliation{Environmental Molecular Sciences Laboratory, Pacific Northwest National Laboratory, Richland, WA, 99352, USA}

\author{Craig M. Brown}
\affiliation{Center for Neutron Research, National Institute of Standards and Technology, Gaithersburg, Maryland 20899, USA}

\author{Anthony K.\ Cheetham}
\affiliation{Materials Department, University of California, Santa Barbara, California 93106, United States}
\altaffiliation{Materials Research Laboratory, University of California, Santa Barbara, California 93106, United States}
\alsoaffiliation{Department of Materials Science and Engineering, National University of
Singapore, 9 Engineering Drive 1, 117575, Singapore}

\author{Pieremanuele Canepa}
\affiliation{Department of Materials Science and Engineering, National University of
Singapore, 9 Engineering Drive 1, 117575, Singapore}
\altaffiliation{Department of Chemical and Biomolecular Engineering, National University of Singapore, 4 Engineering Drive 4, 117585, Singapore}
\email{pcanepa@nus.edu.sg}

\author{Rapha\"{e}le J.\ Cl\'{e}ment}
\affiliation{Materials Department, University of California, Santa Barbara, California 93106, United States}
\altaffiliation{Materials Research Laboratory, University of California, Santa Barbara, California 93106, United States}
\email{rclement@ucsb.edu}

\title{Stacking Faults Assist Lithium-Ion Conduction in a Halide-Based Superionic Conductor}

\begin{document}


\begin{abstract}
In the pursuit of urgently-needed, energy dense solid-state batteries for electric vehicle and portable electronics applications, halide solid electrolytes offer a promising path forward with exceptional compatibility against high-voltage oxide electrodes, tunable ionic conductivities, and facile processing. For this family of compounds, synthesis protocols strongly affect cation site disorder and modulate \ce{Li+} mobility. In this work, we reveal the presence of a high concentration of stacking faults in the superionic conductor \ce{Li3YCl6} and demonstrate a method of controlling its \ce{Li+} conductivity by tuning the defect concentration with synthesis and heat treatments at select temperatures. Leveraging complementary insights from variable temperature synchrotron X-ray diffraction, neutron diffraction, cryogenic transmission electron microscopy, solid-state nuclear magnetic resonance, density functional theory, and electrochemical impedance spectroscopy, we identify the nature of planar defects and the role of nonstoichiometry in lowering \ce{Li+} migration barriers and increasing \ce{Li} site connectivity in mechanochemically-synthesized \ce{Li3YCl6}. We harness paramagnetic relaxation enhancement to enable \textsuperscript{89}Y solid-state NMR, and directly contrast the Y cation site disorder resulting from different preparation methods, demonstrating a potent tool for other researchers studying Y-containing compositions. With heat treatments at temperatures as low as 333~K (60$\degree$C), we decrease the  concentration of planar defects, demonstrating a simple method for tuning the \ce{Li+} conductivity. Findings from this work are expected to be generalizable to other halide solid electrolyte candidates and provide an improved understanding of defect-enabled \ce{Li+} conduction in this class of Li-ion conductors.

\end{abstract}

\section{Introduction}

Achieving greater market penetration for electric vehicles today hinges on gaining the public’s trust in their durability, versatility, and safety. In recent years, Li-ion solid-state batteries (SSBs) with inorganic solid electrolytes (SEs) have gained traction as safer and potentially higher energy density alternatives to the commercial liquid electrolyte (LE) cells. Replacing the combustible organic LE with a nonflammable SE severely lessens the consequences of an internal short-circuit and may also extend the range of operating temperatures for the cell.\cite{janekSolidFutureBattery2016,noiOxideBasedCompositeElectrolytes2018}  Beyond safety, the lack of any liquid component opens the door to bipolar stack SSB architectures that enhance battery module energy density and lower manufacturing costs as cells no longer have to be individually packed to avoid leakage.\cite{jungSolidStateLithium2019, famprikisFundamentalsInorganicSolidstate2019} Notably, Li-ion SEs have transference numbers close to one which could allow for faster charging than LE cells when paired with high ionic conductivities.\cite{janekSolidFutureBattery2016, grohInterfaceInstabilityLiFePO42018, zhangSynthesisCharacterizationArgyrodite2018, griffithNiobiumTungstenOxides2018, adeliBoostingSolidState2019, schlenkerStructureDiffusionPathways2020}

Since a 2018 publication from Asano \emph{et al.} \cite{asanoSolidHalideElectrolytes2018} reported high \ce{Li+} conductivities in \ce{Li3YCl6} (LYC) and \ce{Li3YBr6} (LYB), lithium-containing halide ternaries have emerged as appealing SE candidates owing to their promising room temperature conductivities, strong oxidative stability to high potentials, and hence compatibility with oxide-based cathode materials.\cite{asanoSolidHalideElectrolytes2018, liProgressPerspectivesHalide2020} Sulfide SEs rely on a body-centered cubic (BCC) anion sublattice to ensure low migration barriers and fast \ce{Li+} conduction.\cite{wangDesignPrinciplesSolidstate2015, namFirstPrinciplesDesignHighly2020, parkDesignStrategiesPractical2018}
Oxides require aliovalent doping to achieve appreciable \ce{Li+} conduction through concerted migration, enabled by greater \ce{Li+} concentrations and concomitant occupation of high energy Li  sites.\cite{chenOriginHighLi2015, renOxideElectrolytesLithium2015, heOriginFastIon2017} In contrast, halide SEs exhibit high \ce{Li+} conductivities at stoichiometric Li contents despite having a close packed anion lattice thanks to a combination of intrinsic vacancies and reduced Coulombic interactions between the migrating \ce{Li+} and monovalent anions.\cite{wangLithiumChloridesBromides2019, adelsteinRoleDynamicallyFrustrated2016,zevgolisAlloyingEffectsSuperionic2018} Recent studies have also led to further improvements in conduction through both isovalent\cite{liOriginSuperionicLi3Y1xInxCl62020,wanInitioStudyDefect2021} and aliovalent substitution\cite{liangSiteOccupationTunedSuperionicLixScCl32020,parkHighVoltageSuperionic2020, parkHeatTreatmentProtocol2021, wuStableCathodesolidElectrolyte2021,schlemNa3XEr1XZrxCl62020, kimLithiumYtterbiumBasedHalide2021}, as well as anion mixing.\cite{liuHighIonicConductivity2020} Ternary lithium halides, especially chloride and fluoride-based chemistries, also resist oxidation against high-voltage oxide electrodes on charge due to the strong electronegativity of their anionic species. This could enable the use of high-voltage cathode compositions, such as \ce{LiNi_{0.5}Mn_{1.5}O4}.\cite{wangLithiumChloridesBromides2019,liProgressPerspectivesHalide2020, liSolidElectrolyteKey2015}  

Asano \emph{et al.}\cite{asanoSolidHalideElectrolytes2018} were the first to observe that the ionic conductivity of LYC decreases from $\approx$0.51~mS~cm$^{-1}$ after annealing a mechanochemically synthesized sample. This behavior, especially when compared to the opposite evolution for the LYB analog, is surprising as one would expect the increased crystallinity induced by annealing to favor long range conduction. Later, \emph{ab initio} molecular dynamics (AIMD) calculations reported by Wang \emph{et al.} \cite{wangLithiumChloridesBromides2019} predicted that the LYC structure proposed by Asano \emph{et al.}\cite{asanoSolidHalideElectrolytes2018} should yield conductivities between $\approx$4.5 and $\approx$14~mS~cm$^{-1}$, with an activation energy for \ce{Li+} ion migration of $\approx$0.19$\pm$0.03~eV  (all confidence intervals listed are 1$\sigma$), far smaller than the $\approx$0.40~eV experimentally measured value.  Using X-ray pair distribution function (PDF) analysis, Schlem \emph{et al.}\cite{schlemMechanochemicalSynthesisTool2019} shed some light on the discrepancies between predicted and experimentally-determined conduction properties when they tied synthesis-induced cation site disorder to the conductivity and activation energy of the sample for \ce{Li3ErCl6}, a compound isostructural to LYC. Mechanochemical synthesis induces greater \ce{Er} disorder, generating polyhedral distortions that open bottleneck transition areas and facilitate \ce{Li+} migration. Notably, their PDF analysis of LYC was precluded by fluorescent behavior under X-ray illumination due to their choice of wavelength ($\lambda=$0.5594~\AA) that limited the accuracy of the interpretation of the data. The propensity for disorder on the yttrium sublattice of LYC  was recently demonstrated by Ito \textit{et al.}, who showed using \textit{in situ} X-ray diffraction that two polymorphs of LYC are obtained when heating a mixture of {\ce{LiCl}} and {\ce{YCl3}} powders.{\cite{itoKineticallyStabilizedCation2021}} A metastable, more ionically conductive $\beta$-LYC phase at 450~K but gives way above 600~K to the more stable, commonly reported $\alpha$-LYC polymorph, which retains the identical hexagonal close packed Cl anion sublattice but differs in its arrangement of Y$^{3+}$ cations.

Functional inorganic materials with layered crystal structures are exceptionally prevalent and find use as Li-ion battery cathodes, superconductors, and catalysts. Stacking faults can be pervasive in these materials with synthesis-dependent concentrations but they are rarely investigated in depth as they are notoriously difficult to model and understand, both experimentally and computationally. Unfortunately, this does not allow for an atomic understanding of the dependence of the properties of interest on the material’s defective crystal structure. The presence of stacking faults can be of paramount importance to ionic conduction and redox properties as exemplified by \ce{Ni(OH)2}, a double hydroxide material commonly used as a cathode in Ni hydride metal batteries, where stacking faults greatly ameliorate the electrochemical performance.\cite{tessierStructureNiOH1999} Furthermore, \ce{Na4P2S6} is a Na-ion solid electrolyte where stacking faults generated from a precipitation synthesis stabilize a conductive high temperature phase at room temperature\cite{scholzPhaseFormationSynthetic2021}. A number of other relevant examples also exist in Li- and Na-ion transition metal oxide cathodes\cite{croguennecNatureStackingFaults1997, croguennecHighResolutionElectron1998, boulineauReinvestigationLi2MnO3Structure2009, boulineauStructureLi2MnO3Different2010, boulineauThermalStabilityLi2MnO32012, paulsenLayeredLiMn1999, xiaNaCrO2FundamentallySafe2012, serrano-sevillanoImpactStackingFaults2021, mortemarddeboisseCoulombicSelforderingCharging2019, serrano-sevillanoEnhancedElectrochemicalPerformance2018, clementInsightsNatureEvolution2016, yuElectrochemicalActivitiesLi2MnO32009, liDynamicImagingCrystalline2019, matsunagaDependenceStructuralDefects2016}, Ag nanoparticle catalysts\cite{liSilverCatalystActivated2019}, and ionic conductors\cite{nemudryRoomTemperatureElectrochemical1998, liP2TypeLayered2019}. 

The three-dimensional LYC crystal structure (space group: $P\bar{3}m1$) investigated in this study bears a strong resemblance to the layered structure of \ce{YCl3} (space group: $C2/m$), where both structures exhibit a hexagonal close-packed \ce{Cl-} anion sublattice (ABAB) and a hexagonal \ce{Y^{3+}} arrangement. However, \ce{YCl3} contains (002) planes of octahedral cation voids, in contrast to LYC, which has \ce{Y^{3+}} and \ce{Li+} ions distributed throughout its (001) and (002) planes. Deng \emph{et al.}\cite{dengUnderstandingStructuralElectronic2020} recently demonstrated that \ce{YCl3} is amenable to polymorphism, especially with respect to the related \ce{BiI3}-type structure which shows an alternate ABCABC anion stacking arrangement. According to their calculations, the energy difference between the two structure-types is $\approx$0.1 kJ~mol$^{-1}$ ($\approx$1 meV~atom$^{-1}$), suggesting that alternate stackings, or even stacking faults, are possible in \ce{YCl3}-type structures. Stacking faults have in fact been mentioned for LYC, LYB, and a mixed anion \ce{Li3YCl3Br3} halide composition\cite{asanoSolidHalideElectrolytes2018, liuHighIonicConductivity2020}, although no detailed analysis of these planar defects nor of their impact on ion conduction has been undertaken to date, preventing the establishment of robust design rules for this family of solid electrolytes.

In this study, we investigate the crystal structure of LYC and its conduction properties as a function of synthesis method, using a combination of synchrotron X-ray diffraction (XRD),  neutron diffraction, cryogenic transmission electron microscopy (cryo-TEM), high-resolution \textsuperscript{6,7}Li and \textsuperscript{89}Y nuclear magnetic resonance (NMR), electrochemical impedance spectroscopy (EIS), and density functional theory (DFT) calculations, to reveal a complex metastable defect landscape. We establish a link between these complex structural defects and \ce{Li+} ion conduction. Our in-depth analysis of  X-ray patterns demonstrates the presence of a high concentration of previously unreported stacking faults in mechanochemically-synthesized LYC that generate face-sharing \ce{YCl6^{3-}} octahedra. EIS, pulsed field gradient-NMR (PFG-NMR), and computational investigation of defect structural models reveal that the stacking faults facilitate \ce{Li+} conduction through the structure by lowering \ce{Li+} migration barriers and generating more inter-layer channels for \ce{Li+} transport. However, the defects are shown to be metastable with some disappearing after heat treatment at temperatures as low as 333~K (60$\degree$C). As the defect concentration decreases, \ce{Li+} transport is hindered and the ionic conductivity of the sample decreases. These findings emphasize the importance of defects in promoting long-range \ce{Li+} conduction in halide-type SEs\cite{goraiDevilDefectsElectronic2021}, and suggest that the conduction properties in these structures are enhanced by high defect concentrations associated with metastable states.

\section{Results}
\subsection{Current understanding of the \ce{Li3YCl6} crystal structure}
\label{sec:structdescript}

\begin{figure}[!ht]
\includegraphics[width=3in]{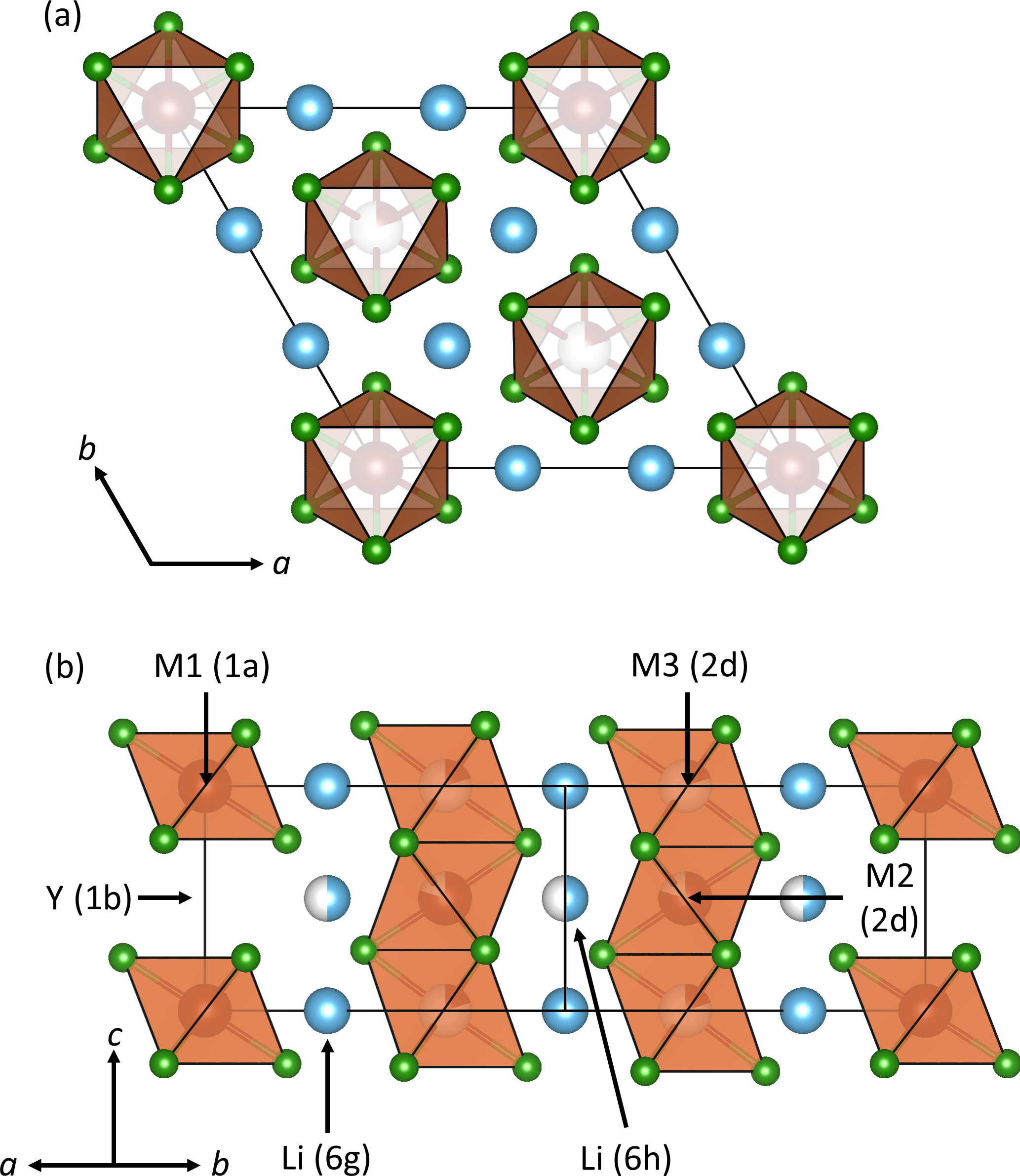}
  \caption{Reported $\mathbf{P\bar{3}m1}$ structure of LYC shown from different angles. (a) structure viewed down the $\mathbf{c}$-axis. (b) Annotated side view of the LYC crystal structure. The structure exhibits full Y occupancy at the M1 (0, 0, 0) position, and split Y occupancy over the M2 (1/3, 2/3, 0.5) and M3 (1/3, 2/3, 0) positions. The Wyckoff positions are denoted in brackets.\cite{schlemInsightsLithiumSubstructure2021} Li positions, as well as the $\mathbf{1b}$ Y site reported by Asano \emph{et al.}\cite{asanoSolidHalideElectrolytes2018}, are also included. }
  \label{fig:average_structure}
\end{figure}

The first report of LYC by Steiner \emph{et al.} \cite{steinerNeueSchnelleIonenleiter1992} proposed LYC crystallizes in an orthorhombic crystal structure. More recently, the trigonal ($P\bar{3}m1$) description of LYC was put forth by Asano \emph{et al.}\cite{asanoSolidHalideElectrolytes2018} and is now deemed to be more appropriate. In the $P\bar{3}m1$ structure, both \ce{Li+} and \ce{Y^{3+}} ions are octahedrally coordinated by \ce{Cl-} ions arranged onto a hexagonal close-packed lattice. However, Rietveld refinements of diffraction patterns of mechanochemically-synthesized LYC by Asano \emph{et al.}\cite{asanoSolidHalideElectrolytes2018} indicate that \ce{Y^{3+}} disorder exists over four Wyckoff sites: $1a$ (0,0,0), $2d$ (1/3, 2/3, 0.5) and (1/3, 2/3, 0), and a $1b$ site (0, 0, 0.5) (Figure~\ref{fig:average_structure}b). This disorder was previously suggested to be the result of the low energy of formation of \ce{Y^{3+}} and \ce{Li+} anti-site defects.\cite{wangLithiumChloridesBromides2019}

An X-ray diffraction study by Schlem \emph{et al.}\cite{schlemMechanochemicalSynthesisTool2019} investigated synthesis-dependent disorder in LYC and a crystallographic isomorph, \ce{Li3ErCl6}. For \ce{Li3ErCl6}, the authors reported that Er atoms fully occupy the $P\bar{3}m1$ $1a$ site denoted as M1, and partially occupy the (1/3, 2/3, 0.5) and (1/3, 2/3, 0) $2d$ sites denoted as M2 and M3, respectively. With only the M1 and M2 sites fully occupied, the structure is referred to as M1 - M2. If the structure has only the M1 and M3 sites fully occupied, it is referred to as M1 - M3. Though X-ray characterization of LYC was hindered by fluorescence for Schlem \emph{et al.} in their study, the insight gained on \ce{Li3ErCl6} proved relevant as the bonding behavior and ionic radii of \ce{Er^{3+}} and \ce{Y^{3+}} (0.890~\AA{} and 0.900~\AA, respectively) are quite similar. They reported that mechanochemical synthesis of \ce{Li3ErCl6} produces a high ionic conductivity sample with an almost entirely M1 - M3 atomic arrangement, while a high temperature annealing synthesis yielded a reduced ionic conductivity sample with an almost entirely M1 - M2 atomic arrangement. A follow-up investigation of LYC by Schlem \emph{et al.}\cite{schlemInsightsLithiumSubstructure2021} using neutron diffraction indicated no appreciable \ce{Li+} occupancy preference between the $P\bar{3}m1$ $6g$ and $6h$ sites (Figure~\ref{fig:average_structure}b), which in turn indicated that the disordered \ce{Y^{3+}} occupancy did not appear to affect the \ce{Li+} substructure. However, we will note that the samples studied by Schlem \emph{et al.}\cite{schlemInsightsLithiumSubstructure2021} were both annealed: one for a week at 823 K (550$\degree$C), and the other first ball milled then annealed for 5 min at 823 K (550$\degree$C). 

Another recent study by Ito \emph{et al.} investigated the synthesis of LYC from its binary precursors, {\ce{LiCl}} and {\ce{YCl3}}, with \emph{in situ} X-ray diffraction.{\cite{itoKineticallyStabilizedCation2021}} Diffraction patterns taken throughout a temperature ramp demonstrated that a new, metastable $\beta$-LYC polymorph ($P\bar{3}c1$) appears above 450~K and is eventually consumed in favor of the previously reported $\alpha$-LYC polymorph ($P\bar{3}m1$) above 600 K. EIS measurements showed that the $\beta$-phase has a higher Li$^+$ conductivity (0.12~mS~cm$^{-1}$) than the $\alpha$-phase (0.014~mS~cm$^{-1}$), which the authors attributed to broadening of the Li$^+$ conduction pathways caused by the higher energy Y$^{3+}$ cation arrangement in the $\beta$-phase. This study showed that the phase transition between the $\beta$ and $\alpha$ polymorphs relies on migration of Y$^{3+}$ along the \textit{c}-axis, which is kinetically hindered at low temperatures and explains the metastability of the $\beta$-phase. Importantly, the authors conclude that the $\beta$-LYC phase is not formed when the precursors are ball milled together as all of the X-ray diffraction peaks reported by Schlem \textit{et al.} could be indexed to $\alpha$-phase reflections.



\subsection{Diffraction characterization of \ce{Li3YCl6}}
\label{sec:neutronresults}
\begin{figure}[ht!]
\includegraphics[width=3in]{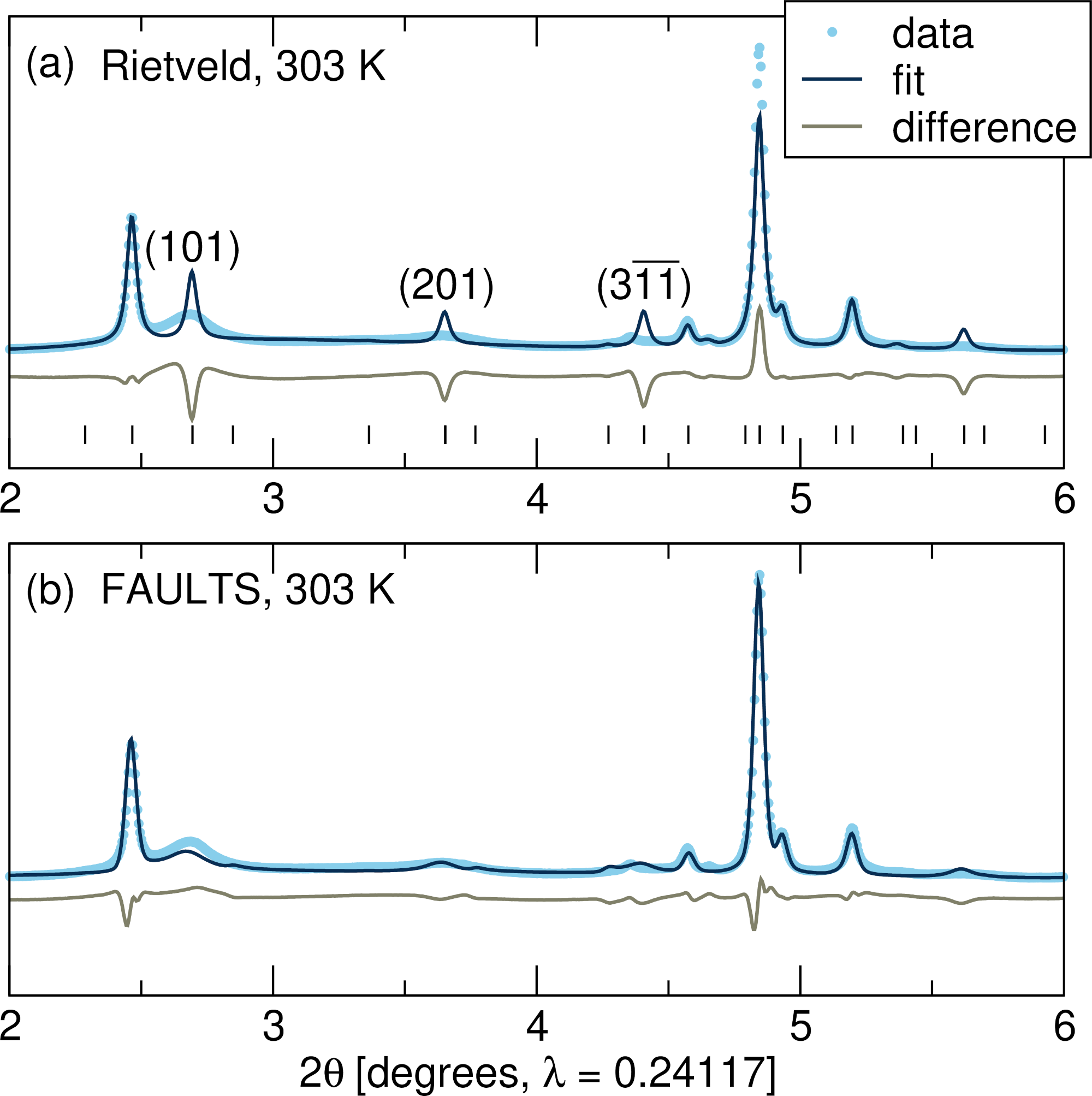}
  \caption{
  Comparison between (a) Rietveld analysis and (b) FAULTS analysis of X-ray diffraction data [17-BM, wavelength = 0.24117~\AA] of ball milled \ce{Li3YCl6} at 303~K. The Rietveld model used a fully occupied M1 site and a fixed-value of 70/30 Y occupancy split over the M2 and M3 sites, respectively, as was reported previously.\cite{schlemInsightsLithiumSubstructure2021}
  }
   \label{fig:fault_compare}
\end{figure}

Ball milled (BM-LYC) and solid-state (SS-LYC) samples of \ce{Li3YCl6} (LYC) were prepared and examined using neutron and X-ray diffraction. The LYC synthesis procedures have been reported elsewhere,\cite{schlemMechanochemicalSynthesisTool2019,schlemInsightsLithiumSubstructure2021} with the distinction that our BM-LYC samples were not annealed after mechanochemical milling, and instead studied at specific temperatures \textit{in situ}. When it comes to structural analysis, the proposed $P\bar{3}m1$ structure is relatively accurate for describing the average structure of LYC. However, as shown in previous work\cite{asanoSolidHalideElectrolytes2018, schlemMechanochemicalSynthesisTool2019}, there is still some ambiguity as to the \ce{Y^{3+}} ordering particularly in BM-LYC. Furthermore, previous models were developed without using X-ray synchrotron diffraction data, which is more sensitive to \ce{Y^{3+}} positions relative to neutron diffraction data. As such, our work utilizing high quality X-ray synchrotron diffraction data reveals important insights into the LYC structure.

\begin{figure*}[ht]
\includegraphics[width=7in]{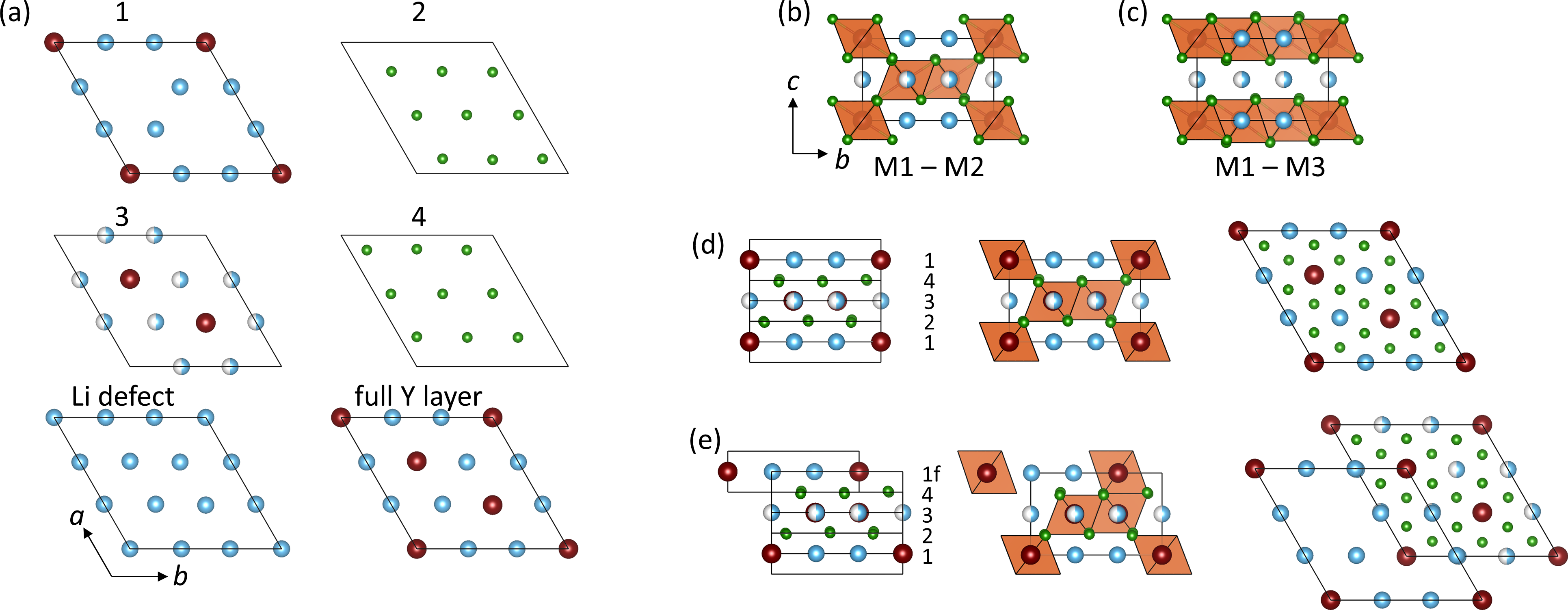}
  \caption{Construction of stacking fault models. (a) individual layers. Atom legend: red for Y; blue for Li; green for Cl. Side on view of the (b) M1 - M2 and (c) M1 - M3 models proposed by Schlem \textit{et al.}\cite{schlemInsightsLithiumSubstructure2021} (d) M1 - M2 layered model that serves as the starting point for the stacking fault model construction (e) illustration of the (1/3, 2/3) layer 1 fault that is responsible for the majority change in the X-ray diffraction pattern. One notes that the \ce{Li+} substructure is essentially unchanged by this fault and the major consequence is the creation of face sharing Y-Cl species along the center of the unit cell.} 
  \label{fig:layers}
\end{figure*}
The importance of synchrotron X-ray data is demonstrated in Figure~\ref{fig:fault_compare}, with the presence of significant broadening and even disappearance of reflections associated with specific $hkl$ planes. This broadening and loss of intensity has previously been attributed to decreased crystallinity of the materials due to ball milling\cite{schlemMechanochemicalSynthesisTool2019}. However, when certain $hkl$ reflections in a diffraction pattern are systematically impacted, such as broadened or even eliminated, this is usually indicative of stacking faults within a sample.\cite{serrano-sevillanoEnhancedElectrochemicalPerformance2018, mortemarddeboisseCoulombicSelforderingCharging2019} Antisite disorder can also be ruled out as these defects would not induce selective broadening. Stacking faults are notorious for complicating the intensities and peak shapes of Bragg reflections in powder diffraction patterns\cite{berlinerEffectStackingFaults1986}, making Rietveld analysis unreliable. 

Using the $P\bar{3}m1$ structural model proposed by Schlem \emph{et al.}\cite{schlemInsightsLithiumSubstructure2021} as a reference, the most impacted $hkl$ reflections in the BM-LYC patterns are the $(101)$, $(201)$, and ($3\bar{1}\bar{1}$). These reflections, as we discuss below, are predominantly caused by planes of Y atoms. As shown in Figure~\ref{fig:fault_compare}a, the peaks in the experimental data are much broader (or absent) relative to the peak intensity expected from the Rietveld model. Figure~\ref{fig:fault_compare}b shows a fit using a stacking fault model, which we believe to be more appropriate for describing BM-LYC. Though there are likely many other structural intricacies present in disordered compounds like LYC (regardless of preparation), our results suggest that LYC prepared \emph{via} ball milling is particularly susceptible to stacking faults. Cryo-TEM on the BM-LYC was attempted to obtain direct evidence for the presence of stacking faults, but the sample's susceptibility to beam damage precluded any observation of the sample in its pristine state (see Figure~S1a,b).

Overall, we find synchrotron X-ray diffraction data to be more useful than neutron diffraction data for the analysis of the structure of LYC samples. Though neutron diffraction was expected to provide additional insight, analysis of the neutron data, as illustrated in Figure~S2, is complicated by the large chloride neutron scattering signal relative to the other elements (combined scattering lengths of Cl $\approx$9.577~fm~$\times$~6 = 57.462~fm,  Li: $\approx$-1.90~fm~$\times$3 = $-5.7$~fm, and Y $\approx$7.75~fm). Figure~S2 illustrates how, even when using drastically different stacking fault models to describe the 303~K BM-LYC diffraction data, the difference between the fits is minimal. This tolerance to changing stacking fault models when fitting the neutron data is a result of the stacking faults involving predominantly Y atoms, as detailed below, which are better observed with X-rays as opposed to neutrons.  

To construct Y atom centric stacking fault models, we used the proposed $P\bar{3}m1$ LYC structure as a starting point.\cite{schlemInsightsLithiumSubstructure2021} This structure captures allowed $hkl$s relatively well (Figure~\ref{fig:fault_compare}a), and indicates that the $hkl$ reflections associated with Y atoms are the most affected by stacking faults. We constructed a layer-by-layer model of the 3D structure that was initially "ordered", but into which \ce{Y^{3+}} disorder could be introduced with relative ease. Our starting stacking sequence reproduced the proposed M1 - M2 structure for \ce{Li3YCl6} from Schlem \emph{et al.}\cite{schlemMechanochemicalSynthesisTool2019} As can be seen in Figure~\ref{fig:layers}a, the layers used to construct the stacking fault model include two Cl layers, a Li defect layer, as well as three Y/Li layers that have Y atoms located at either the (0,0), (1/3, 2/3) and (2/3, 1/3), or all of those sites. Figure~S3 illustrates in the form of a flow chart how the fault model refines the percent occurrence of each layer as the layer sequence deviates from the one present in the M1 - M2 structure, all the while retaining trigonal symmetry. Figure~\ref{fig:layers}b illustrates the M1 - M2 model from Schlem \emph{et al.}\cite{schlemMechanochemicalSynthesisTool2019}, and Figure~\ref{fig:layers}d illustrates the layer sequence used to build the starting M1 - M2 structure.

As can be seen in Figure~\ref{fig:layers}a, layer 1 has Y atoms only at the (0, 0) position. The fault that causes the broadening/loss of the (101), (201), and (3$\bar{1}\bar{1}$) peaks is comprised of the two equally likely (1/3, 2/3) and (2/3, 1/3) faults of layer 1 (as illustrated in Figure~\ref{fig:layers}e).  This fault explains why the proposed average model by Schlem \textit{et al.}\cite{schlemMechanochemicalSynthesisTool2019, schlemInsightsLithiumSubstructure2021} is relatively accurate in describing the LYC structure, as it introduces face sharing columns of Y-Cl octahedra at the (1/3, 2/3) and (2/3, 1/3) positions that propagate along the \textit{c}-axis. In the $P\bar{3}m1$ structure, this fault is approximated by allowing \ce{Y^{3+}} to occupy M3 sites (Figure~\ref{fig:average_structure}). However, the $P\bar{3}m1$ model is inappropriate as it splits \ce{Y^{3+}} occupancy over the M2 and M3 sites, with a combined crystallographic site occupancy of one (to charge balance the model). The issue is that if the layer 1 fault exists, and one attempts to explain the structure using the $P\bar{3}m1$ model, then the occupancy of M3 should be shared with the M1 sites, not the M2 sites. Furthermore, though the fault model rationalizes why M3 occupancy exists in the $P\bar{3}m1$ model, it is unclear why there would be M2 sites without full occupancy. This discrepancy can be resolved by introducing Li defect layers with \ce{Li+} ions on all possible sites that remove \ce{Y^{3+}} occupancy from the M2 sites.

\begin{figure}[H]
\includegraphics[scale=1.2]{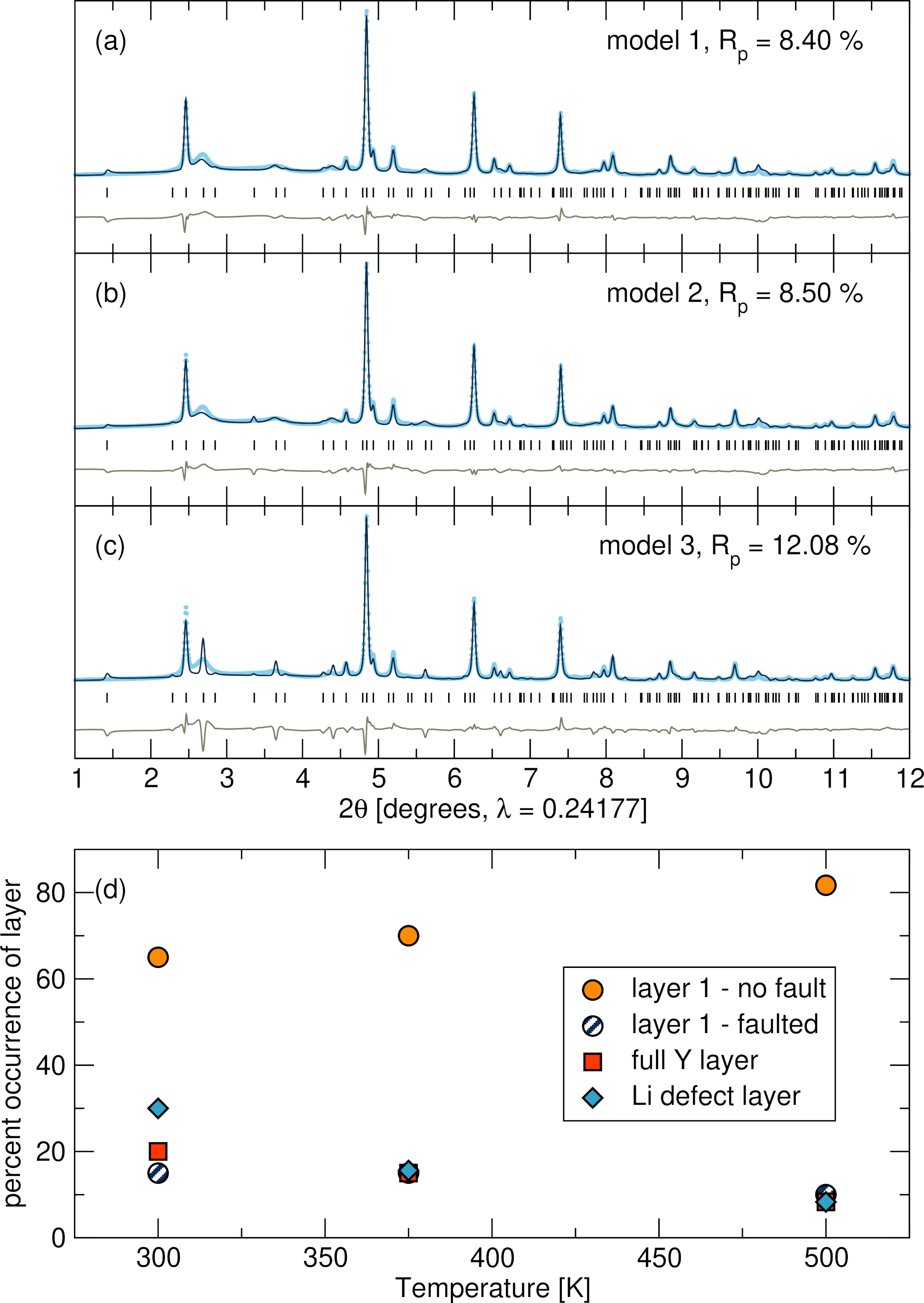}
  \caption{Results from FAULTS refinements of synchrotron X-ray diffraction data [17-bm, APS] of the BM-LYC sample at 303~K with and without certain faults/defect layers included. (a) Model 1 fit. This model exhibits layer 1 Y-fault $\mathbf{[}$(1/3, 2/3) and (2/3, 1/3) shift$\mathbf{]}$, Li defect layer, and full Y layer occurrences of 15\%, 30\%, and 20\%, respectively. The layer unit cell dimensions were refined to \textit{a} = 11.20675(4), \textit{c} = 3.02971(9) \AA{} where a layer unit cell is a three layer slab comprising a Li/Y layer bound by two Cl layers. (b) Model 2 fit. This model has a layer 1 Y-fault $\mathbf{[}$(1/3, 2/3) and (2/3, 1/3) shift$\mathbf{]}$ occurrence of 22\% and a Li defect layer occurrence of 17 percent. The layer unit cell was refined to \textit{a} = 11.20337(3), \textit{c} = 3.02351(7) \AA. (c) Model 3 fit. This fit illustrates the effect of using only the full Y layer and Li defect layer. Note how, by not using any layer 1 fault [(1/3, 2/3) and (2/3, 1/3) shift], the peak broadening beneath 5 \textdegree2$\theta$ is not appropriately captured. The full Y layer occurrence is at 25\%, and the Li-defect layer at 50\%. The layer unit cell dimensions were refined to \textit{a} = 11.20640(5) and \textit{c} = 3.01845(2) \AA, respectively. \hl{For clarity, the 2$\theta$ range plotted in (a), (b), and (c) extends from 1 to 12\textdegree, but the refinements were performed between 1 and 15\textdegree. The tick marks shown correspond to expected \textit{hkl} reflections for the proposed average structure (space group $P\bar{3}m1$, as shown in Figure 2(a)).} (d) Percent occurrence for layers in model 1 as temperature is increased, as determined from FAULTS refinement of X-ray diffraction data [17-bm, APS]. Error estimates are one standard deviation.} 
  \label{fig:model_compare}
\end{figure}

Figure~\ref{fig:model_compare} shows comparative fits of the BM-LYC diffraction data at 303 K with various stacking fault models. Two models of faults and defect layers were found to best fit the diffraction data of BM-LYC, denoted as model 1 and model 2 hereafter. Model 1 is the preferred of these two models as it better explains the temperature evolution of the diffraction data of the BM-LYC sample. While model 1 and 2 are almost identical, model 1 also includes the "full Y layer" as a possible occurrence. The full Y layer is used to emulate the Y/Li layers that exist in the M1 - M3 structure (Figure~\ref{fig:layers}c) reported by Schlem \emph{et al.}\cite{schlemMechanochemicalSynthesisTool2019} We note here that the inclusion of full Y layers in model 1 requires an equivalent amount of Li defect layers so as to maintain 3-1-6 stoichiometry and to charge balance the full Y layer. Inclusion of Li defect layers significantly improves the fits (see Figure~S4). The presence of Li defect layers also explains some of the structural changes observed \emph{via} NMR upon heating the BM-LYC, as discussed in Section~\ref{sec:NMRresults}.

Figure~\ref{fig:model_compare}a shows the model 1 fit, which has a 15\% occurrence of the layer 1 fault [(1/3, 2/3) and (2/3, 1/3) shift], as well as a 20\% occurrence of the full Y layer. Interestingly, the fit is improved by including more Li defect layers than is necessary to charge balance the full Y layers. Figure~\ref{fig:model_compare}b shows the model 2 fit, which has more layer 1 faults (22\% occurrence) and fewer Li defect layers (17\%) relative to model 1. Figure~\ref{fig:model_compare}c (model 3) illustrates how inappropriate it is to exclusively rely on the M1 - M3 layer and Li defect layers to simulate the pattern, as doing so does not lead to the observed peak broadening. 

The BM-LYC structure was characterized via diffraction at various temperatures between 303~K to 500~K. The evolution of the percent occurrence of the different layers accounted for in model 1 (at 303~K, 375~K, and 500~K) is plotted in Figure~\ref{fig:model_compare}d. As can be seen from the plot, all faults/defects decrease as temperature increases, with the overall structure becoming more like the M1 - M2 model. The Li defect layer used for modeling accounts for two types of Li layers - the Li layer required to charge balance the existence of the full-Y layer, as well as any Li rich defects. Though these two types of Li layers might differ in Li content/arrangement, the use of one type of Li layer is an effective approximation for both layers because Li has minimal X-ray scattering contribution relative to Y and Cl atoms. The main purpose of including a Li defect layer is to account for a lack of Y scattering density in the X-ray diffraction data. At 303~K, the percent occurrence of the Li defect layer is greater than the necessary amount to charge balance the full Y layer. This indicates that the BM-LYC compound is off-stoichiometric and best thought of as \ce{Li_{3+3x}Y_{1-x}Cl_6}. At 375~K and 500~K, the Li defect layer occurrence approximately matches the full Y layer, indicating that the compound is closer to the expected \ce{Li3YCl6} stoichiometry. In contrast, a refinement using model 2 indicates that Li defect layers still exist at 500~K. One may understand now why model 2 is less preferred, as these types of defects are likely to disappear completely at elevated temperature. 


A BM-LYC composition of \ce{Li_{3+3x}Y_{1-x}Cl_6} at 303~K is difficult to ascertain from refinements alone. However, further evidence for off-stoichiometry in the room temperature sample comes from the emergence of an \ce{LiCl} phase upon exposure to elevated temperatures, as discussed in Section~\ref{sec:NMRresults}. If \ce{LiCl} were to appear from the decomposition of LYC, we would expect to also observe diffraction peaks corresponding to \ce{YCl3}, which we do not. Hence, \ce{LiCl} is likely precipitating out of the LYC phase, which suggests the presence of Li-rich regions in the LYC structure.

We now turn our attention to the SS-LYC sample. While the SS-LYC is prepared using a traditional solid-state method and involved repeated regrinding and reheating steps, this sample contains multiple phases. This can be inferred from the diffraction patterns shown in Figures~S5 and S6. Figure~S5 illustrates how the SS-LYC pattern obtained upon ramping up to 500~K shares features that are identical in shape and intensity to those observed in the BM-LYC sample held at 500~K for 50 mins (both patterns were collected under identical conditions). Such shared features between the two types of LYC samples are only seen in the high temperature data sets. Unfortunately, simply subtracting the BM-500 K 50 min pattern from the SS–LYC 500~K pattern does not produce a pattern of sufficient quality for Rietveld refinement. As can be seen from Figure~S6, the LiCl peak present in the SS-LYC 300 K pattern decreases in intensity after heating to 500~K, implying that \ce{LiCl} (and other components of the multi-phasic SS-LYC sample) is still reacting upon heating to 500~K. The presence of the $\beta$-LYC polymorph was ruled out based on a Rietveld refinement presented in Figure~S7. Its absence agrees well with the results from Ito \textit{et al.} as our week-long synthesis was conducted at 823~K, far outside of the reported temperature stability window for the $\beta$-phase.{\cite{itoKineticallyStabilizedCation2021}}   Electron diffraction measurements obtained \textit{via} cryo-TEM (Figure~S1c,d) confirm the presence of domains containing stacking faults in SS-LYC. Streaks in an electron diffraction pattern have been widely reported as indications of stacking faults{\cite{whelanElectronDiffractionCrystals1957, bianStackingFaultsTheir2001, yamaneEffectsStackingFault2009}} and provide unquestionable evidence that the LYC layered structure is capable of experiencing these planar defects.

Overall, as the SS-LYC sample is multi-phasic with one of those constituent phases containing stacking faults (the BM-LYC-like features), we are unable to develop reliable structural models to fit its diffraction patterns. 

\subsection{Computational analysis of Li-Y orderings and stacking faults}

With compelling evidence for the presence of stacking faults from the diffraction analysis, we turned to theoretical simulations to better understand the propensity for disorder on the Y/Li lattices in LYC. Specifically, computing the energies of different LYC structural models with DFT offers insight into the most probable Li-Y arrangements in the structure. 

Starting from the LYC unit cell (with space group $P\bar{3}m1$) as reported by Asano \emph{et al.}\cite{asanoSolidHalideElectrolytes2018} which includes four possible Y positions (see Section~\ref{sec:structdescript}), 237 distinct Li/Y arrangements were enumerated.\cite{ongPythonMaterialsGenomics2013} Of these, 76 are models with 3 formula units (\ce{Li9Y3Cl18}), while 161 are supercells including up to 6 formula units (\ce{Li18Y6Cl36}). In parallel, another set of structures was enumerated starting from specific stacking fault models identified by X-ray diffraction (see Section~\ref{sec:neutronresults}), comprising up to 15 formula units (\ce{Li45Y15Cl90}) of LYC. In these stacking fault models, Li appears with fractional occupation, which we address using the same enumeration procedure as mentioned above. In total, 285 unique orderings were computed, of which 48 are symmetrically distinct stacking fault models, using DFT within the SCAN meta-GGA approximation.  This level of theory appears adequate for the simulation of LYC and its structural features (see Section~S3 of the Supporting Information).

The relative stability of each LYC ordering was assessed in terms of its decomposition into the \ce{LiCl} and \ce{YCl3} binary compounds that share the same composition line. 
Notably, all the LYC orderings (including the stacking fault models) considered here always decompose into \ce{LiCl} and \ce{YCl3}, in agreement with previous investigations.\cite{wangLithiumChloridesBromides2019} The relative instability of the different LYC models is quantified by the energy above the convex hull, i.e., the propensity of LYC to decompose into \ce{LiCl} and \ce{YCl3}. Structural models with energies above the convex hull $\leq$30~meV~atom$^{-1}$ are likely to be stabilized by entropy contributions at $\approx$298~K (or higher temperature), and hence may be accessible \emph{via} high-temperature or high-energy mechanochemical synthesis protocols. 

\begin{figure}
\includegraphics[width=\columnwidth]{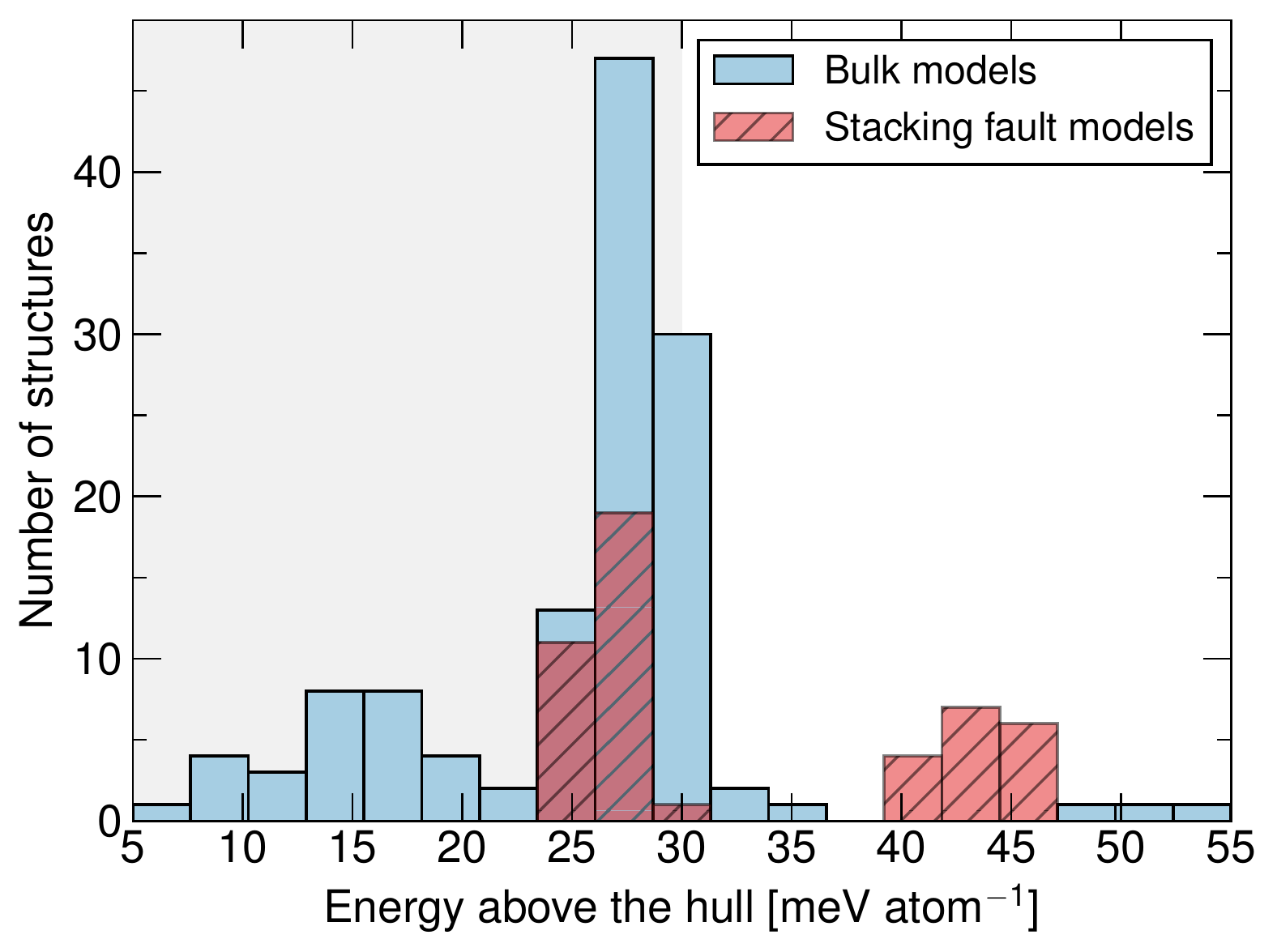}
  \caption{
  Distribution of predicted Li/Y orderings in \ce{Li3YCl6} vs.\ their relative thermodynamic stability, expressed as energy above the hull. Some of the 237 bulk orderings are shown with blue bars, whereas the 48 stacking faults by the red-hatched bars. Energetically accessible orderings below the 30~meV~atom$^{-1}$ threshold are enclosed in the gray shaded area. The figure displays only orderings up to 55 meV~atom$^{-1}$.
  } 
  \label{fig:theorystacking}
\end{figure}

Figure~\ref{fig:theorystacking} plots the distribution of energy above the hull of the bulk orderings and stacking fault orderings considered in this work, which were computed at 0~K, excluding any entropic or $pV$ effects not explicitly accounted for in these simulations. We find that a significant number of structural orderings (i.e., 142) fall below the 30~meV~atom$^{-1}$ threshold, including $\approx$111 bulk orderings and $\approx$31 stacking faults models. This result suggests a rich, experimentally-accessible configurational landscape for the arrangement of cation species, as well as a complex LYC structure containing both bulk-like and stacking fault features. The thermodynamic stability of each structural model in Figure~\ref{fig:theorystacking} is directly influenced by the distribution of Y/Li cations among the octahedral sites formed by the Cl anion framework. Only specific Y/Li orderings minimize the electrostatic repulsion between \ce{Y^{3+}} and \ce{Li+} ions, and correspond to those with a low energy above the convex hull. Notably, the low energy structures in Figure~\ref{fig:theorystacking} feature \ch{Li} face-sharing chains, in excellent agreement with previous observations by  Schlem \textit{et al.}\cite{schlemInsightsLithiumSubstructure2021} In general, the Y ordering in LYC has a significant impact on the relative stability of each structural model, while the energetics of our models are less dependent on the Li ordering. This observation is in agreement with the minimal 6g/6h site preference previously observed by Schlem \emph{et al.}\cite{schlemInsightsLithiumSubstructure2021}, our DFT results reinforce the idea that Li is expected to be extremely mobile in LYC. 

Our calculations on bulk structures of \ch{Li3YCl6} indicate that low energy structures, i.e., with energies above the convex hull below ~25~meV~atom$^{-1}$, never display facesharing \ce{YCl6^{3-}} octahedra. Yet, some higher energy structures (both stacking fault and bulk models) below the 30~meV~atom$^{-1}$ threshold do include facesharing \ce{YCl6^{3-}} octahedra and are likely accessible \emph{via} mechanochemical synthesis. Therefore, the formation of stacking faults among other stable bulk configurations in LYC is expected after ball milling.

\subsection{Probing Li and Y local environments with NMR}
\label{sec:NMRresults}

\begin{figure*}[ht!]
\includegraphics[width=7in]{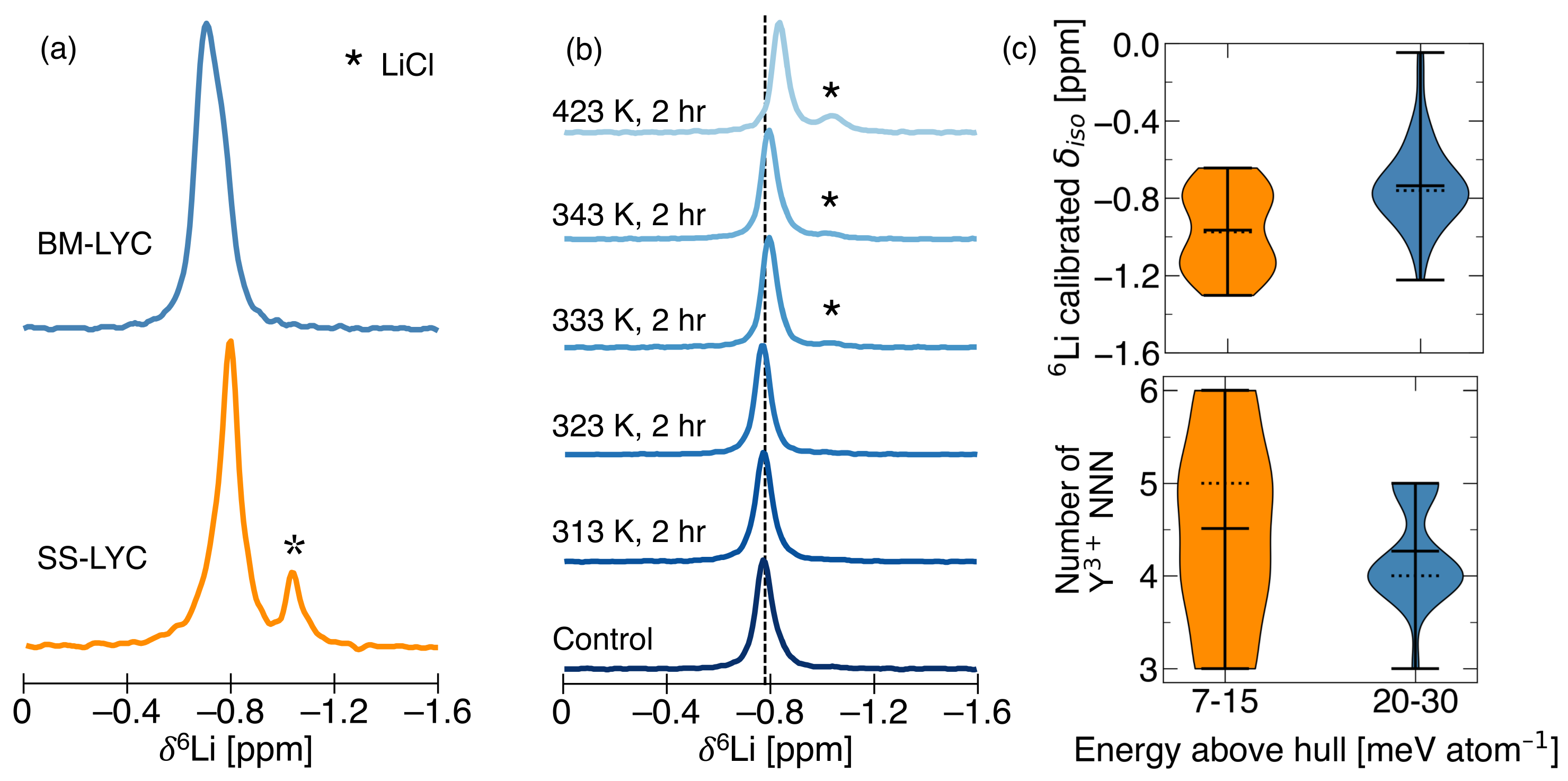}

  \caption{
  (a) $^{6}$Li NMR spectra for as-prepared BM-LYC and SS-LYC. (b) $^{6}$Li NMR spectra collected on a series of BM-LYC samples heat treated at various temperatures from room temperature to 423~K (150$\degree$C). Besides the control sample which was only sealed in a capillary, all samples were heated in a sealed capillary for a period of $\approx$2~hours. \ce{LiCl} is labeled with an asterisk (*). All spectra were acquired at 18.8~T with a spinning speed of 25~kHz and a set temperature of 298~K. (c) Distribution of DFT-calculated $^{6}$Li isotropic chemical shifts and number of \ce{Y^{3+}} next nearest neighbors ($NNN$) for Li atoms in enumerated LYC orderings grouped according to their energy above the convex hull from Figure~{\ref{fig:theorystacking}}. The lowest energy LYC ordering was $\approx$7~meV~atom$^{-1}$ above the hull. Computed structures included some instances of \ce{YCl_{6}^{3-}} face-sharing octahedra. The solid and dotted horizontal bars inside the shaded areas represent the mean and median values, respectively.
  \label{fig:6Li_NMR_castep}
  }
\end{figure*}

Having recognized the presence of stacking faults through diffraction and DFT simulations, the Li and Y local environments in the BM-LYC and SS-LYC samples were investigated using solid-state NMR spectroscopy. Notably, NMR is a local structure technique and is therefore a powerful probe of the phases present in samples regardless of their degree of crystallinity. Since the presence of (1/3, 2/3) Y layer faults yields different local environments for the Y atoms near the faulted layer (see Figure~\ref{fig:layers}e), $^{89}$Y NMR can be used to differentiate between LYC phases with varying amounts of such faults. Simultaneously, $^{6}$Li NMR provides high-resolution insight into the Li substructure.

The differences in the Li substructures between samples prepared \emph{via} ball milling or solid-state synthesis were first explored, and the $^{6}$Li NMR spectra of the two samples are presented in Figure~\ref{fig:6Li_NMR_castep}a. The main $^{6}$Li signal for SS-LYC is centered at $\approx$--0.82~ppm and clearly resolved from its \ce{LiCl} impurity peak at $\approx$--1.05~ppm. A single resonance is observed in the spectrum collected on BM-LYC and is centered at  $\approx$--0.70~ppm, which is offset to more positive ppm values relative to the solid-state sample. The BM-LYC $^{6}$Li NMR lineshape is broader and less symmetrical than that of its solid-state counterpart, suggesting that it is composed of several overlapping signals with very similar resonant frequencies due to a distribution of Li environments in the more disordered BM-LYC structure.  

The decrease in the amount of stacking faults and defect layers upon heating, as observed \emph{via} diffraction (see Figure~\ref{fig:model_compare}d), warrants an NMR-based investigation of the effect of heat treatments on the Li substructure. Capillaries containing BM-LYC powder samples were sealed, held for $\approx$2~hours at various temperatures up to 423~K (150$\degree$~C), and air quenched to characterize the temperature-induced structural evolution. The $^{6}$Li NMR spectra obtained at room temperature on these samples are plotted in Figure~\ref{fig:6Li_NMR_castep}b. A control sample that was sealed in a capillary but not heat treated was also measured to ensure that any effects from flame-sealing could also be accounted for. With increasing temperature, the $^{6}$Li lineshape shifts to more negative ppm values and approaches the isotropic shift of the SS-LYC sample. Starting at 333~K (60$\degree$~C), a \ch{LiCl} component appears and continues to grow at higher temperatures, suggesting that the evolution of the BM-LYC disordered structure is linked to the presence of \ce{LiCl}.

DFT calculations of $^{6}$Li NMR parameters were performed on enumerated bulk structures considered for the analysis of structural energetics (Figure~\ref{fig:theorystacking}) to elucidate the origin of the temperature-dependent $^{6}$Li shift of BM-LYC. Figure~\ref{fig:6Li_NMR_castep}c depicts the distributions of calculated $^{6}$Li shift values ($^{6}$Li shift calibration provided in Section~S5) and of the number of \ce{Y^{3+}} next nearest neighbors ($NNN$) to Li atoms in selected structural models, which are grouped according to their energy above the convex hull. Two groupings are considered: the first comprises the low energy structures with energies ranging between $\approx$7 and 15 meV~atom$^{-1}$ (see Figure~\ref{fig:EIS_fig}), while the higher energy bin corresponds to structures with energies between $\approx$20-30~meV~atom$^{-1}$, and contains instances of \ce{YCl_{6}^{3-}} face-sharing octahedra. Full stacking fault structures, which contain more formula units per unit cell, were not calculated due to prohibitively high computational cost. While structures in the 20 meV~atom$^{-1}$ - 30 meV~atom$^{-1}$ range (blue bin in Figure~\ref{fig:6Li_NMR_castep}c) have shifts that are mostly centered near $\approx$--0.75~ppm, $^{6}$Li shifts for the lower energy structures are found to have more negative chemical shifts, as evidenced by the inferior mean and median values. When the number of \ce{Y^{3+}} $NNN$ to each Li atom is considered, it can be seen that the more negative chemical shift of the low energy structures is correlated with a greater number of surrounding \ce{Y^{3+}} ions.

Before any exposure to elevated temperatures, the BM-LYC sample can be thought of as consisting of structures in the 20 meV~atom$^{-1}$ - 30 meV~atom$^{-1}$ range, which include instances of face-sharing \ce{YCl6^{3-}} octahedra. However, the heat treatment provides enough thermal energy to enable structural rearrangements leading to lower energy configurations ($\approx$7 meV~atom$^{-1}$ - 15 meV~atom$^{-1}$), as evidenced by the shifting of the average $^{6}$Li resonance. This yields a more negative $^{6}$Li chemical shift for the SS-LYC sample annealed at 823~K (550~$^{\circ}$C) for multiple days.  330~K (60~$^{\circ}$C) appears to be a threshold temperature for the stability of BM-LYC: above this temperature, the structure begins to evolve as demonstrated by the shift of the $^{6}$Li resonances to more negative ppm values. Attributing the decreasing $^{6}$Li chemical shift in BM-LYC to an increase in the number of \ce{Y^{3+}} $NNN$ to each Li atom is consistent with the elimination of Li-only defects, as observed from diffraction.

To facilitate the acquisition of $^{89}$Y NMR data, \ch{Sm^{3+}} was doped into the LYC structure, whereby the paramagnetic \ch{Sm^{3+}} species replace some of the diamagnetic \ce{Y^{3+}} ions in the nominal \ch{Li_3Y_{0.95}Sm_{0.05}Cl_6} composition (LYC-Sm) and trigger paramagnetic relaxation enhancements (PRE) that drastically shorten the $^{89}$Y longitudinal (T1) relaxation times. The shorter $^{89}$Y NMR signal lifetime leads to a significant reduction in the recycle delay used between NMR scans, which enables the acquisition of $^{89}$Y spectra with acceptable signal-to-noise ratio.\cite{grey89YMagicAngle1990} Notably, doping of paramagnetic \ch{Sm^{3+}} into the LYC structure does not result in significant crystallographic differences (see XRD patterns in Figure~S9) due to the similar ionic sizes of \ce{Y^{3+}} (0.900~\AA) and \ce{Sm^{3+}} (0.958~\AA) in octahedral environments.\cite{shannonRevisedValuesEffective1970}

\begin{figure}[ht!]
\includegraphics[scale=0.5]{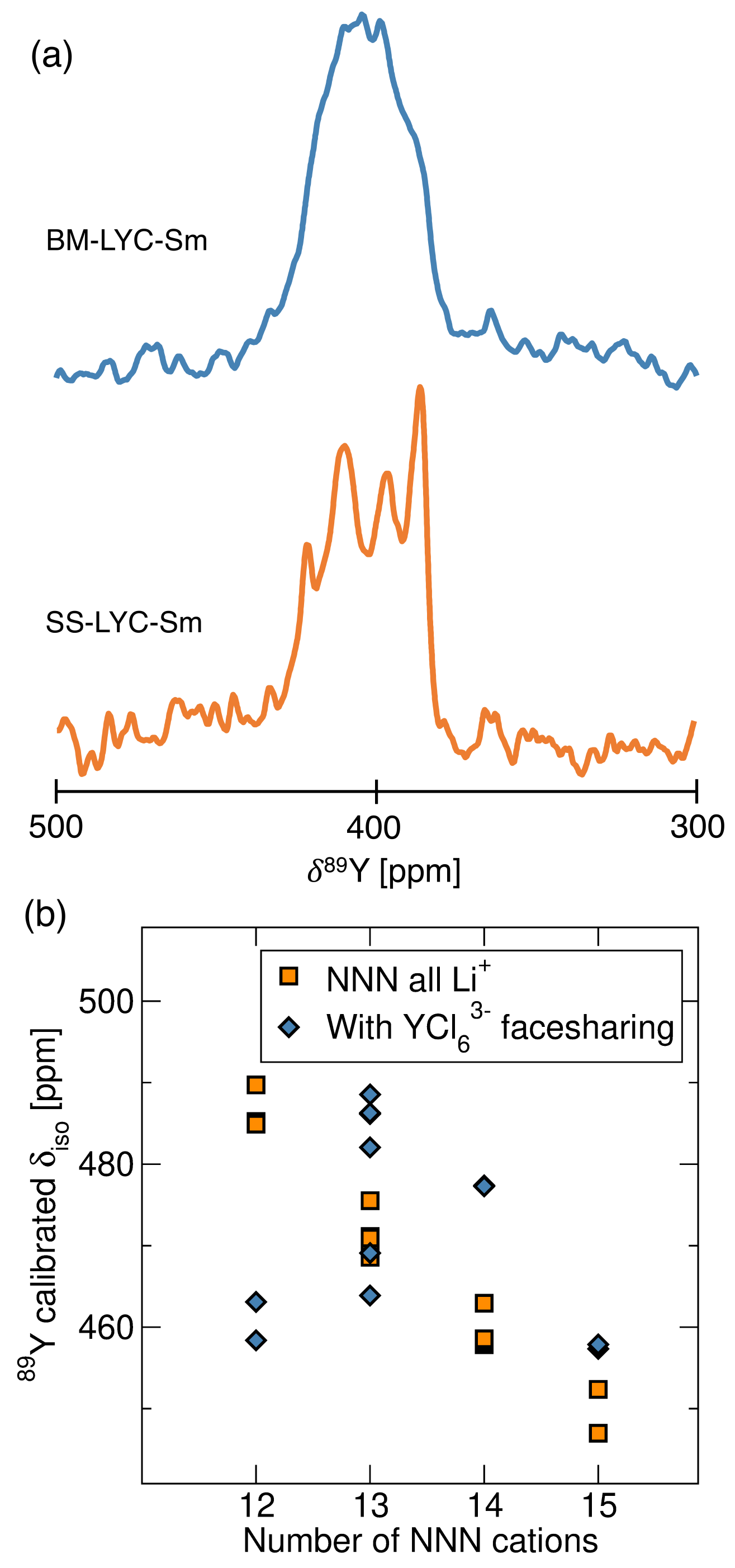}

  \caption{
  (a) $^{89}$Y resonances for the Sm-doped LYC samples, BM-LYC-Sm and SS-LYC-Sm. All spectra were acquired at 18.8~T and a 10~kHz MAS rate with a set temperature of 298~K. (b) Isotropic $^{89}$Y chemical shifts plotted as a function of the number of next nearest neighbor ($NNN$) cations for each corresponding \ce{Y^{3+}}. The shifts were computed with DFT for bulk orderings within 25.5 meV~atom$^{-1}$ from the convex hull in Figure~\ref{fig:theorystacking}.
  \label{fig:89Y_BMvSS-LYC}
  }
\end{figure}

Figure~\ref{fig:89Y_BMvSS-LYC}a shows the $^{89}$Y NMR spectra collected on the Sm-doped ball milled and solid-state LYC samples acquired at 298~K. The presence of 0.05 \ce{Sm^{3+}} does not induce a paramagnetic shift for $^{6}$Li or $^{89}$Y, but slightly broadens the $^{6}$Li resonance relative to the undoped sample (see Figures~S10, S11). The $^{89}$Y NMR signals observed in the spectra collected on the SS- and BM-LYC-Sm samples fall within the same $\approx$390~ppm to $\approx$460~ppm chemical shift range, but have dissimilar lineshapes. The SS-LYC-Sm $^{89}$Y spectrum is made up of four, clearly resolved resonances, centered at 386 ppm, 396 ppm, 410 ppm, and 421 ppm, respectively. A fit of the spectrum results in four peaks corresponding to 19\%, 27\%, 43\%, and 11\% of the total integrated $^{89}$Y signal intensity (see Figure~S12). In contrast, a single broad signal is observed in the BM-LYC-Sm $^{89}$Y spectrum, which is presumably composed of closely spaced and overlapping  resonances as expected for increased disorder and a distribution of Y environments in the material. Due to the difficulty in deconvoluting individual resonances, the BM-LYC $^{89}$Y spectrum cannot be fitted reliably.

To assist the interpretation of the $^{89}$Y NMR spectra, DFT calculations of $^{89}$Y isotropic shifts in enumerated bulk LYC orderings with and without face-sharing between \ce{YCl_{6}^{3-}} octahedra were conducted (see details in Computational Methods section). Results from these calculations are plotted in Figure~\ref{fig:89Y_BMvSS-LYC}b. A group of 8 structures was computed (6 \ce{Li9Y3Cl18} unit cells and 2 \ce{Li18Y6Cl36} supercells) with every structure lying within $\approx$25.5~meV~atom$^{-1}$ from the hull. The calculated isotropic shifts were converted to experimentally-relevant values using a calibration curve constructed from a range of reported $^{89}$Y shifts for Y-containing compounds (see Section~S10 for calibration curve data and additional details). Structures are classified into two categories according to whether or not they contain face-sharing \ce{YCl_{6}^{3–}} octahedra. For the structures that do not, all the $NNN$ of \ce{Y^{3+}} cations are octahedral \ce{Li+}, mostly edge and corner sharing \emph{via} Cl anions. The $^{89}$Y chemical shifts corresponding to these environments fall into four clear ranges according to the number of \ce{Li+} $NNN$, with more \ce{Li+} neighbors resulting in a more negative shift value, which is consistent with previous reports on Y-containing alumino-silicate glasses.\cite{jaworskiScandiumYttriumEnvironments2017} The four computed chemical shift bins are reminiscent of the four signals observed in the $^{89}$Y NMR spectrum of the SS-LYC-Sm compound in Figure~\ref{fig:89Y_BMvSS-LYC}a, suggesting that the $NNN$ of \ce{Y^{3+}} cations in the solid-state sample are almost all \ce{Li+}. The ppm differences between the average shift of each bin are calculated to be 15~ppm, 12~ppm, and 10~ppm, respectively, and agree remarkably well with the separations between the fitted peaks for the SS-LYC-Sm $^{89}$Y spectrum of 10~ppm, 14~ppm, and 11~ppm. We note that comparisons between the computed and experimental shifts are based on relative ppm differences, rather than absolute values, as the calibration curves derived from first-principles can lead to a chemical shift offset.\cite{readerCationDisorderPyrochlore2009} 

Inclusion of structures containing face-sharing \ce{YCl_{6}^{3–}} octahedra breaks the aforementioned binning as it leads to a broad chemical shift range for a given number of $NNN$ \ce{Li+}. While some spectral broadening is to be expected for a mechanochemically synthesized sample due to a wider range of possible bond lengths and angles in the strained, defective structure, the lack of clear clusters of computed $^{89}$Y chemical shifts is consistent with the broad NMR spectrum for BM-LYC-Sm. This observation is in agreement with the presence of face-sharing \ce{YCl_{6}^{3–}} octahedra in BM-LYC ---a key feature of our proposed stacking fault model.

\subsection{Evaluation of \ce{Li+} ion conduction properties}

\begin{figure*}[ht!]
\includegraphics[width=6in]{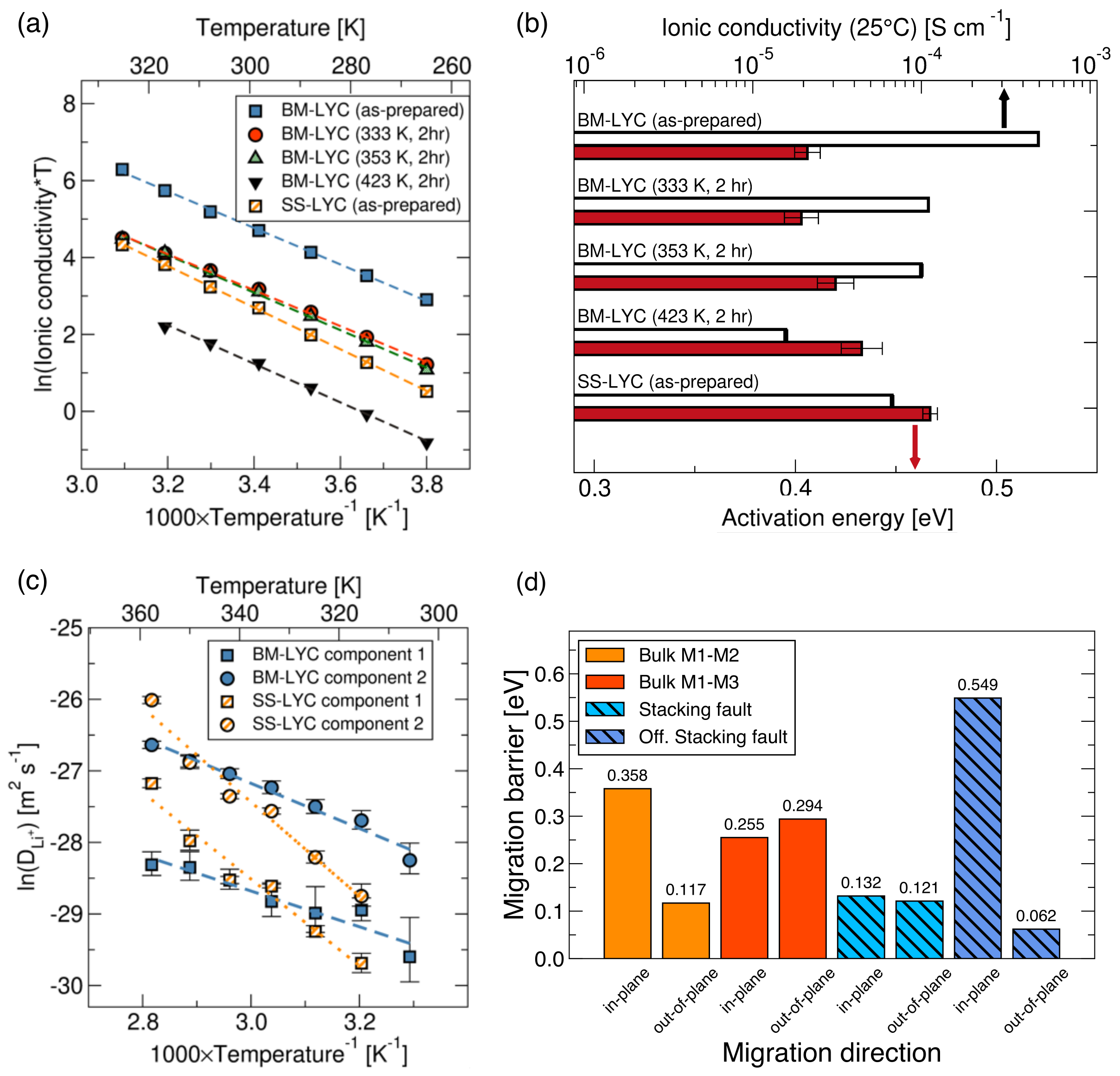}
  \caption{
  (a) Arrhenius behavior of EIS-measured ionic conductivity for as-prepared BM-LYC and heat treated samples (b) Room temperature ionic conductivity and activation energy for \ce{Li+} long-range diffusion for various LYC samples (c) Variable temperature PFG-NMR measurements on as-prepared BM- and SS-LYC samples. All measurements were conducted at 7.05~T. (d) Predicted migration barriers in (eV) for \ce{Li+} ions in different model structures: bulk M1 - M2 (orange bars), bulk M1 - M3 (red bars), stacking fault (light blue), and off-stoichiometric stacking fault (blue). The uncertainty of the migration barriers is $\pm$0.06~eV.\cite{chenIonicTransportPotential2019} All error estimates are for one standard deviation. 
  \label{fig:EIS_fig}
  }
\end{figure*}

The \ce{Li+} ion conductivities and activation energies of a series of LYC samples were measured with electrochemical impedance spectroscopy (EIS). Representative Nyquist plots and fits are provided in Figure~S14. Fits of the EIS data were performed using an equivalent circuit comprising a parallel resistor and a constant phase element (CPE), to capture the combined grain and grain boundary conduction, in series with a CPE to capture the electrodes' response. Measured conductivities therefore do not deconvolute grain and grain boundary contributions, providing instead a total conductivity through the LYC pellet. 

Plots of the temperature dependence of the ionic conductivity, as well as of the room temperature conductivities and activation energies, are provided in Fig~\ref{fig:EIS_fig}a-b. The measured ionic conductivity at 298~K (25$\degree$C) (of 0.49~mS~cm$^{-1}$ and 0.067~mS~cm$^{-1}$) and activation energy (of 0.41$\pm$0.006~eV and 0.47$\pm$0.004~eV) (all confidence intervals listed are 1$\sigma$) for the as-prepared BM-LYC and SS-LYC samples, respectively, are in good agreement with published values from Asano \emph{et al.}\cite{asanoSolidHalideElectrolytes2018} and Schlem \emph{et al.}\cite{schlemMechanochemicalSynthesisTool2019} The influence of metastable defects and of the Li substructure on \ce{Li+} conduction in BM-LYC is probed by measuring the EIS response of samples heat treated under analogous conditions to the samples considered in Figure~\ref{fig:6Li_NMR_castep}b. As shown in Figure~\ref{fig:EIS_fig}b, the heat treatments at 333~K, 353~K, and 423~K decrease the room temperature ionic conductivity of the samples relative to the as-prepared compound. At temperatures higher than 333~K (60$\degree$C), \ce{Li+} diffusion barriers increase, suggesting that the evolution of the BM-LYC structure above 333~K affects \ce{Li+} conduction pathways. While previous experiments by Schlem \emph{et al.}\cite{schlemMechanochemicalSynthesisTool2019} tested 5 min and 1 hour of annealing at 823~K (550$\degree$C), the temperatures tested here are far lower, emphasizing the facile tunability of BM-LYC's \ce{Li+} conductivity upon exposure to moderate temperatures. 


Variable temperature $^{7}$Li PFG-NMR measurements on BM- and SS-LYC are shown in Figure~\ref{fig:EIS_fig}c. PFG-NMR tracks the self-diffusion of  $^{7}$Li nuclei ($D_{Li^+}$) through a material by applying a magnetic field gradient across the height of the sample, giving spins a spatial encoding, and measuring their NMR signal after a time ($\Delta$) during which the \ce{Li+} ions diffuse. Relative to EIS, PFG-NMR probes short range diffusion on the nm-$\mu$m length scale and can access intra-grain conduction properties.\cite{harmLessonLearnedNMR2019} For BM-LYC, diffusion constants could be measured down to $\approx$304~K on account of its superior diffusion at relatively low temperatures. In both LYC samples, two diffusing components are observed that can be attributed to \ce{Li+} ion conduction within LYC particles. All measured self-diffusion constants are tabulated in Table~S5.
 
 The derived activation energies for ion migration in BM-LYC are 0.25$\pm$0.01~eV and 0.18$\pm$0.03~eV  (all confidence intervals listed are 1$\sigma$) for components 1 and 2, respectively. For SS-LYC, these are 0.57$\pm$0.09~eV and 0.48$\pm$0.1~eV. The measured activation energies of SS-LYC are comparable to the 0.47~eV EIS-measured value, while the values for BM-LYC are much lower than the measured 0.41~eV. Because the temperature range probed in our PFG-NMR measurements extends up to 363~K (90$\degree$C), which is higher than the critical temperature of 333~K (60$\degree$C) above which the long-range \ce{Li+} conductivity of BM-LYC decreases, the measured activation energies for BM-LYC should be considered as lower bounds.  
 
 Variable diffusion time ($\Delta$) PFG-NMR measurements were also conducted to check for impeded transport across grain boundaries, and the results are displayed in Figure~S15. The diffusion length for a migrating spin is given by $r=(6D\Delta)^{1/2}$. Hence, monitoring self-diffusion constants as a function of the applied $\Delta$, and comparing the diffusion length with the average grain size of a sample can provide information about the relative ease of intra-grain vs.\ inter-grain diffusion processes.
 
 No change in the $D_{Li^+}$ is observed for SS-LYC as the $\Delta$ time is increased, while BM-LYC $D_{Li^+}$ values monotonically decrease with increasing $\Delta$. This behavior is consistent with impeded conduction across grain boundaries, i.e., as the diffusion length scale exceeds the BM-LYC average grain size, but cannot be deconvoluted from potential effects from heating at 363~K (90$\degree$C). In fact, the larger errors in the $D_{Li^+}$ values for the BM-LYC sample may be caused by structural evolution from heating, or by a distribution of spin-spin relaxation times ($T_{2}$) due to the presence of a wide range of $^{7}$Li environments. SEM images as well as $D_{Li^+}$ values and length scales can be found in the supporting information (Table~S6, Figure~S16).


To better understand the impact of stacking faults on the \ce{Li+} conduction pathways, bond valence maps were calculated on four model LYC structures with varying Li/Y site occupations and defect layers. Bond valence maps are 3D potential maps for the diffusing \ce{Li+} ions. They are determined by the bond valence site energy (BVSE) method and provide insight into how \ce{Li+} ions migrate through a crystal structure.\cite{adamsHighPowerLithium2011,chenBondSoftnessSensitive2017} Of the four structures analyzed, two are free of stacking faults and correspond to an M1 - M2 configuration (i.e., equal Y occupation of $1a$ and z = 0.5 $2d$ sites) and an M1 - M3 configuration (i.e., Y only in 1a and z = 0 $2d$ sites), respectively.\cite{schlemMechanochemicalSynthesisTool2019} In the M1 - M2 model (Figure~S17), \ce{Li+} conduction occurs primarily along the \textit{c}-axis, emphasizing facile \ce{Li+} migration along O\textsubscript{h}-O\textsubscript{h} paths between face-sharing octahedra, as predicted by Wang \emph{et al}.\cite{wangLithiumChloridesBromides2019} The M1 - M3 structure (Figure~S18), while retaining the preferential \textit{c}-axis diffusion, displays increased \textit{ab}-plane connectivity as the Y-free z = 0.5 layer provides greatly decreased Coulombic repulsion between \ce{Li+} and \ce{Y^{3+}} ions. Bond valence maps for two additional faulted structural models obtained from fits of the X-ray patterns in Figure~\ref{fig:model_compare} were also computed and reveal new site linkages. The first of these structures contains only (1/3, 2/3) Y layer faults and is illustrated in Figure~S19. While facesharing \ce{YCl_{6}^{3–}} octahedra decrease the \ce{Li} site connectivity in their vicinity due to strong Coulombic repulsion with migrating \ce{Li+} ions, they lead to regions of enhanced Li site connectivity elsewhere through the appearance of “loop” connectivity features, which bridge four neighboring Li sites across the \textit{ab} plane and \textit{c} axis. An additional structure that includes both Y (1/3, 2/3) faults, as well as the Y-free Li defect layers (Figure~S20), illustrates how both faults work in tandem to further increase Li site connectivity, as the loop features appear alongside the linked Li sites of the Li defect layer along the ab-plane. Altogether, bond valence maps of faulted structures highlight how the presence of defects can locally increase the degree of Li site connectivity, which provides an intuitive explanation for the improvement in \ce{Li+} conductivity upon ball milling.


First-principles calculations were carried out to assess the transport of \ce{Li+} ions in bulk LYC models and in the vicinity of planar defects.\cite{wangLithiumChloridesBromides2019} \ce{Li+} migration barriers are influenced by the presence of Y and other \ce{Li+} in the vicinity of the migrating ion. Given the large number of possible Li/Y orderings and stacking fault models for LYC (see Figure~\ref{fig:theorystacking}), the number of migration barriers to be considered can easily grow exponentially. Here, we only compute the \ch{Li+} migration barriers in the same structural models used for bond valence mapping and along different crystallographic directions. The results of these calculations are presented in Figure~\ref{fig:EIS_fig}d. The migration barriers were computed with DFT (at the SCAN level of theory) and using the nudged elastic band (NEB) method (see Computational Methods section).\cite{henkelmanClimbingImageNudged2000} The accuracy of the migration barriers in Figure~\ref{fig:EIS_fig}d is estimated to be $\pm$0.06~eV,\cite{chenIonicTransportPotential2019} which approximately matches one order of magnitude in diffusivity.

Macroscopic Li transport is enabled by a percolation network of \ch{Li+} migration pathways in the particles of LYC. Two distinct migration paths can result in Li percolation in LYC, nominally, \emph{i)} 2D \ce{Li+} migration in the $ab$ plane between octahedral sites bridged by a tetrahedral site ---these pathways are referred as in-plane. \emph{ii)} \ce{Li+} migration perpendicular to the 2D layers of Li (along the crystallographic direction $c$) and termed out-of-plane (see Figure~\ref{fig:average_structure}). This anisotropic diffusion mechanism has been previously reported for LYC and other layered \ce{Li3MX6} compositions.\cite{asanoSolidHalideElectrolytes2018,xuInfluenceAnionCharge2019, wangLithiumChloridesBromides2019}

Stoichiometric LYC contains intrinsic vacancies on the Li sublattice, which actively participate in \ce{Li+} migration. In Figure~\ref{fig:EIS_fig}d, the \ce{Li+} migration barriers for the M1 - M2 and M1 - M3  bulk models as well as the stoichiometric stacking fault model use the existing vacancy network without removing a Li atom. In contrast, for the off-stoichiometric fault model, all of the Li sites in the Li-only defect layer are filled so a vacancy was introduced in the structure to enable ion migration.

The results shown in Figure~\ref{fig:EIS_fig}d indicate that the lowest energy M1 - M2 bulk ordering ($\approx$7.56~meV~atom$^{-1}$ above the hull) leads to in-plane and out-of-plane computed barriers of 0.358~eV and 0.117~eV, respectively. The M1 - M3 bulk ordering ($\approx$9.8~meV~atom$^{-1}$ above the hull), results in in-plane and out-of-plane barriers of 0.255~eV and 0.294~eV, respectively. The stacking fault model with the lowest energy above the convex hull ($\approx$24.10~meV~atom$^{-1}$ above the hull) with the \ce{Li3YCl6} stoichiometry leads to in-plane and out-of-plane barriers of 0.132~eV and 0.121~eV, respectively. The model for an off-stoichiometric stacking fault results in in-plane and out-of-plane barriers of 0.549~eV and 0.62~eV, respectively.

From Figure~\ref{fig:EIS_fig}d, one notices a large variability in the migration barriers derived for the different bulk or stacking fault models. In general, the migration barriers of in-plane pathways are  larger than those of out-of-plane pathways. An exception to this trend is the set of barriers computed for the M1 - M3 model. The lowest computed barrier is that of Li migrating out-of-plane in the off-stoichiometric fault. In fact, the out-of-plane \ce{Li+} migration barriers are consistently lower in the stacking fault models than in the bulk models (i.e., M1 - M2 and M1 - M3). These results suggest that stacking faults are key to facile \ch{Li} transport in LYC.

\section{Discussion}

The results presented above emphasize the complexity of the defect landscape in LYC and its role in promoting \ce{Li+} conduction through the structure. While previous literature had exclusively reported Y$^{3+}$ site-disorder, our work demonstrates that mechanochemical synthesis of LYC can cause the coexistence of stacking faults and Li-only defect layers, as evidenced by a combination of high resolution synchrotron X-ray diffraction data, NMR, and DFT. These defects are metastable, allowing for a highly tunable conductivity. When ball milled samples are exposed to temperatures above 333~K (60$\degree$C), the defect concentration decreases and \ce{Li+} conduction is reduced.

Examining the \ce{Li+} transport properties of LYC with the combination of EIS and PFG-NMR probes  migration of \ce{Li+}. The two diffusing components observed with PFG-NMR (Figure~\ref{fig:EIS_fig}c) likely correspond to in-/out-of-plane diffusion. Since $c$-axis conduction usually has a lower migration barrier, component 1 is assigned to diffusion along the $c$-axis and component 2 to $ab$-plane diffusion. For the ball milled sample, both components exhibit much lower activation energies than that of bulk \ce{Li+} conduction measured by EIS (0.25$\pm$0.01~eV and 0.18$\pm$0.03~eV vs. 0.41$\pm$0.006 eV), reflecting that \ce{Li+} conduction within individual particles is more facile than across grain boundaries and likely enhanced by the high concentration of defects within each particle. NEB calculations informed by bond valence maps show that stacking faults and Li defect layers generate additional site linkages with lower migration barriers and facilitate \ce{Li+} ion transport. These defects create areas of dense \ce{Y^{3+}} cation distribution (i.e., Y face-sharing moieties) and other areas of sparse \ce{Y^{3+}} distribution, where migrating \ce{Li+} ions experience less repulsive forces from nearby \ce{Y^{3+}} ions. Among the \ce{Li3MX6} family of compounds, sparser \ce{M^{3+}} distributions have been reported to improve \ce{Li+} conduction and are a critical parameter for designing better halide Li-ion solid electrolytes.\cite{kimMaterialDesignStrategy2021} While intra-particle conduction is facilitated by defects, grain boundaries still hinder long-range \ce{Li+} conduction, as evidenced by both the decreasing \ce{Li+} self-diffusion constants at longer $\Delta$ values from PFG-NMR, and the high activation energy from EIS for bulk migration.

Previous computational investigations of \ce{Li+} migration in LYC have mostly relied on \emph{ab initio} molecular dynamic simulations (AIMD),\cite{wangLithiumChloridesBromides2019} and are often limited to a single Li/Y ordering in the bulk structure and specific direction for \ce{Li+} migration. Undoubtedly, AIMD provides appealing results that are directly ``comparable''  with experimental EIS measurements. However, our NEB calculations are advantageous as they isolate favorable mechanisms of \ce{Li+} migration and link \ce{Li+} transport directly to the underlying structural motif --- information often lost in the collective nature of AIMD simulations. In general, our computed migration barriers agree well with previous computational data.\cite{wangLithiumChloridesBromides2019,kimMaterialDesignStrategy2021} Using AIMD on a M1 - M2-bulk configuration, Wang \emph{et al.}\cite{wangLithiumChloridesBromides2019} derived an average migration barrier of $\approx$0.19$\pm$0.03~eV, which is in agreement with our findings for out-of-plane diffusion. In contrast, they reported a much smaller barrier ($\approx$0.23$\pm$0.06~eV) (all confidence intervals listed are 1$\sigma$) for the in-plane \ce{Li+} migration pathway ($\approx$0.358~eV in this work). Furthermore, Wang \emph{et al.}\cite{wangLithiumChloridesBromides2019} showed that introducing one and three anti-site defects between \ce{Li+} and \ce{Y^{3+}} yields an average activation energy of $0.28\pm0.03$~eV and $0.31\pm0.03$~eV, respectively. Similarly, Kim \emph{et al.}\cite{kimMaterialDesignStrategy2021}, who included explicit van der Waals corrections in their AIMD simulation, obtained a larger average \ce{Li+} migration activation energy of $\approx$0.370~eV. While the existing data is in agreement with our results of Figure~\ref{fig:EIS_fig}d, by explicitly considering the occurrence of stacking faults in the LYC structure, our simulations go beyond the analysis of \ce{Li+} migration in pristine bulk structures, which cannot reflect the defect-rich mechanochemically synthesized structures. Further, the use of local experimental and computational tools in this work is justified by the strong dependence of the LYC short-range structure and \ce{Li+} mobility on the synthesis protocol. In turn, our results enable us to provide microscopic insights into previous computational and experimental observations on halide solid electrolytes.

While Schlem \emph{et al.}\cite{schlemMechanochemicalSynthesisTool2019} had reported significant changes in the site disorder of LYC for very short times of annealing at 823~K (550$\degree$C), here, we show that the defect concentration begins to decrease at $\approx$333~K (60$\degree$C), leading to a lower ionic conductivity as determined from EIS. Variable temperature synchrotron XRD and NMR measurements on heat treated samples indicate the elimination of \ce{Li+} defect layers and concurrent generation of LiCl. LiCl, being far less conductive, likely confines some of the \ce{Li+} and prevents it from contributing to bulk conduction. The non-stoichiometry of LYC implied by the presence of Li-only defect layers is likely due to the fact that \ce{YCl3} sticks to the agate mortar used for precursor mixing, a phenomenon also reported by Schlem \emph{et al.}\cite{schlemMechanochemicalSynthesisTool2019} The emergence of LiCl upon heating has also been reported for \ce{Li3YbCl6}\cite{kimLithiumYtterbiumBasedHalide2021}, suggesting that non-stoichiometry and Li-defect layers could be present in other \ce{Li3MCl6} compositions. Stacking faults also appear to be eliminated at higher temperatures, although the exact mechanism is yet unknown. We speculate that stacking faults could be eliminated either through Y$^{3+}$ migration, individual layer slippage that removes facesharing between \ce{YCl_{6}^{3–}} octahedra, or through grain growth, which consumes smaller, faulted particles in favor of new unfaulted shells on larger pre-existing particles. Answering this question will be the subject of future investigations. 

Although this is the first experimental investigation of stacking faults among \ce{Li3MX6} structures, it is likely that other kindred compositions are prone to similar stacking faults. The small energy difference between alternate stacking sequences for \ce{YCl3}\cite{dengUnderstandingStructuralElectronic2020}, and its similarity to the LYC crystal structure, serve as a possible explanation for the susceptibility of this class of materials to stacking faults. Given the similarities between the crystal structures of \ce{YCl3} and other \ce{MCl3} compounds (\ce{M}=\ce{Yb}, \ce{Er}), as well as their tendency to form trigonal \ce{Li3MCl6}, it is likely that \ce{Li3YbCl6} and \ce{Li3ErCl6} are susceptible to stacking faults when made \emph{via} mechanochemical synthesis. Furthermore, evidence from Deng \emph{et al.}\cite{dengUnderstandingStructuralElectronic2020} suggests that layered structures are favorable for larger halide anions as well and that alternate stacking sequences remain relatively close in energies. As such, \ce{Li3MX6} compositions where \ce{X}=\ce{Br} or \ce{I} may also be prone to stacking faults. While these have yet to be reported, a computational investigation by Xu \emph{et al.}\cite{xuInfluenceAnionCharge2019} on the possible stacking sequences of \ce{Li3LaI6} has demonstrated that alternate stacking sequences result in energetically similar structures. Therefore, stacking fault formation is likely and could occur during mechanochemical synthesis. Future work on the family of \ce{Li3MX6} compounds should take care to ensure that potential stacking faults are accounted for in crystallographic analysis, and seek to better understand stacking fault susceptibility according to the identity of the alkali ion, anion, metal species.

In order to be used in commercially viable devices, solid electrolytes must maintain their conduction properties over a range of temperatures. The metastability of the planar defects that facilitate \ce{Li+} conduction in LYC limits its range of applications as its ionic conductivity decreases markedly above 333~K (60$\degree$C). Further investigation into strategies for stabilizing these defects and a better understanding of the factors that control their creation during synthesis is required to commercialize halide solid electrolytes. 

\section{Conclusion}
A high concentration of stacking faults and other defects are identified in mechanochemically-synthesized LYC and tied to its \ce{Li+} conductive properties. Harnessing synchrotron X-ray diffraction, we have shown that that the as-made BM-LYC exhibits stacking faults that generate face-sharing \ce{YCl6^{3–}} octahedra. The BM-LYC is also suspected to have lithium rich layers within the material, suggesting a slight degree of off-stoichiometry. Cryo-TEM observations of domains containing stacking faults in the SS-LYC sample provide unquestionable evidence of the LYC layered structure's susceptibility to these planar defects. DFT calculations confirm that the observed defects are within the realm of entropically-stabilized configurations at room temperature. Variable temperature diffraction, as well as $^6$Li and $^{89}$Y solid-state NMR reveal that heat treatments, even at temperatures as low as 333~K (60$\degree$C), reduce the concentration of defects in the material. Our PFG-NMR and EIS measurements, probing local and macroscopic \ce{Li+} conduction, show that the defect concentration is intimately tied to the ionic conductivity of LYC; decreasing defect concentration \emph{via} low temperature heat treatments is a means of tuning the ionic conductivity. Bond valence maps reveal new \ce{Li} site linkages in the vicinity of planar defects as the locally sparse \ce{Y^{3+}} distribution decreases the repulsive forces experienced by migrating \ce{Li+}. Furthermore, NEB calculations confirm that these defects facilitate ionic conduction by lowering the \ce{Li+} migration energy barrier. While these results constitute the first investigation of stacking faults among the ternary Li metal halides, we expect other kindred compositions may suffer from these faults. As such, future studies of these compounds must account for their presence and seek to better understand their formation.

\section{Experimental}
\subsection{Material synthesis}
All samples were synthesized under an inert Ar atmosphere. LiCl (99+\%, Aldrich Chemical company) was dried in a vacuum oven at 225 $\degree$C under dynamic vacuum for 24 hours to remove residual moisture before being transferred into the glovebox. \ce{YCl3} (99.99\% trace metals basis, anhydrous, Sigma-Aldrich) was used as received. Precursors were hand-ground together with an agate mortar and pestle for 20 minutes with a 10\% weight excess of \ce{YCl3} to compensate for preferential adhesion of \ce{YCl3} to the surface of the mortar, in accordance with previous reports.\cite{schlemMechanochemicalSynthesisTool2019} \ce{SmCl3} was used to dope LYC with \ce{Sm^{3}+} on the \ce{Y^{3}+} site to trigger a paramagnetic relaxation enhancement (PRE) that made \textsuperscript{89}Y NMR more tractable, analogous to the procedure used by Grey \emph{et al.} when studying pyrocholores.\cite{grey89YMagicAngle1990} When doping with Sm, a stoichiometric amount of \ce{SmCl3} (99.9\% trace metals basis, anhydrous, Sigma-Aldrich) was added to powder mixture before grinding. For the ball mill synthesis, 1.6 g of hand-mixed powder was loaded into a $\approx$45~ml \ce{ZrO2} jar along with seven 10~mm and fourteen 5~mm Y-stabilized zirconia spherical grinding media and sealed under Ar. The powder was milled at 8.3 Hz (500 rpm) for a total of 297 cycles of 5 min milling followed by 15 min rest with a high energy planetary ball mill (Retsch PM 200). Every 99 cycles, the jar was opened under Ar and caked-on powder was scraped from the edges with a spatula. For the solid-state synthesis, $\approx$~1.8 g of hand-mixed powder was pressed into four roughly equivalent 6 mm diameter pellets and flame sealed in a fused quartz tube (10 cm long, 13 mm inner diameter) under vacuum. Ahead of the sealing, the fused quartz tube was dried in a vacuum oven at 225 $\degree$C under dynamic vacuum for 12 hours. The pellets were annealed at 550 $\degree$C for a total of 6 days in a tube furnace (Thermo Scientific Lindberg Blue M). To ensure a homogeneous final product, the pellets were removed from the tube, ground in an agate mortar, and repelletized after 2 days and 4 days of annealing.

To investigate structural evolution as a function of temperature, ball milled LYC was flame sealed in capillaries and placed in a temperature-controlled environmental chamber (Tenney TJR-A-F4T). Capillaries were held there for 2 hours before being removed and opened in the glovebox. Samples that were never exposed to elevated temperatures but were sealed in capillaries served as controls to prove the methodology did not lead to air or moisture exposure.

\subsection{Diffraction}
X-ray diffraction measurements were performed on $\approx$40 mg and $\approx$50 mg samples of ball milled and solid-state prepared \ce{Li3YCl6} powders, respectively, at beamline 17-BM at the Advanced Photon Source at Argonne National Laboratory. The  temperature  of  the  capillary  samples  was  achieved  using  an  Oxford Cryosystems Cryostream 800. Scattered  intensity was measured by a PerkinElmer amorphous-Si flat panel detector. The wavelength for the measurements was 0.24117 Å. Samples were measured first at 303~K, then at specific temperatures on heating up to 500~K. The heating rate was $\approx$ 1 degree K per minute.  The ball milled sample was also measured at 500~K after being held at that temperature for 50 minutes. Data was analyzed using the FAULTS program\cite{casas-cabanasFAULTSProgramRefinement2016} (based on DIFFaX\cite{treacyGeneralRecursionMethod1991}) within the FullProf software suite.\cite{rodriguez-carvajalRecentAdvancesMagnetic1993} The TOPAS software suite was used to show how the Rietveld refinement treatment of these data sets is inappropriate.\cite{coelhoTOPASTOPASAcademicOptimization2018}

Neutron diffraction measurements were performed on a $\approx$1.2 g and $\approx$1.3 g samples of ball milled and solid-state prepared \ce{Li3YCl6} powders, respectively, at the National Institute of Standards and Technology Center for Neutron Research (NCNR). Data was collected at the high-resolution neutron powder diffractometer, BT-1, utilizing a Cu(311) monochromator with an in-pile 60’ collimator, corresponding to a neutron wavelength of 1.5400~\AA. Each sample was loaded into a vanadium sample can in a He environment glove box and sealed with a soldered lead o‑ring onto copper heating block. After mounting each sample onto a bottom-loaded closed cycle refrigerator (CCR), the same was measured at 303~K for a sufficient time so as to have appropriate data statistics. Data was analyzed using the FAULTS program\cite{casas-cabanasFAULTSProgramRefinement2016} (based on DIFFaX\cite{treacyGeneralRecursionMethod1991}) within the FullProf software suite.\cite{rodriguez-carvajalRecentAdvancesMagnetic1993}

\subsection{Transmission electron microscopy}
LYC samples were studied by cryo-TEM on a
Titan 80-300\textsuperscript{TM} scanning/transmission electron microscope (S/TEM) operated at 300 kV. Particles were dispersed onto a TEM lacey carbon grid inside an Ar-filled glovebox. The specimen was then taken out into an air-tight container and immediately plunged into liquid nitrogen followed by transferring the specimen onto a pre-cooled cryo-TEM holder (Elsa, Gatan, USA) using a cryotransfer station to ensure the entire process occurred under a cryogenic environment, which keeps the specimen in its native state. Then, the cryo-TEM holder was inserted into the TEM column for characterization at low temperature (100 K).

\subsection{NMR}
All one dimensional spectra were acquired at 18.8~T (800 MHz for \textsuperscript{1}H) on a Bruker Ultrashield Plus standard bore magnet equipped with an Avance III console. \textsuperscript{6}Li measurements were done on a 2.5~mm HX MAS probe with 2.5~mm single cap zirconia rotors packed and closed with a Vespel cap under Ar with a PTFE spacer between the sample and cap to further protect the sample from air exposure. A flow of \ce{N2} gas at 2000~L~hr\textsuperscript{–1} was used to control the rotor temperature and protect the sample from moisture contamination. Rotors were spun at 25~kHz using dry nitrogen and data were obtained using a rotor synchronized spin-echo pulse sequence (90$\degree$-TR-180$\degree$-TR-ACQ). 90$\degree$ and 180$\degree$ flip angles of 3.125~$\mu$s and 6.25~$\mu$s, respectively, at 200~W were used with a 60~s recycle delay between scans. \textsuperscript{6}Li chemical shifts were referenced to a 1 mol/L \ce{LiCl} liquid solution at 0~ppm. Pulse lengths were calibrated on the same solution. All spectra were processed with Topspin 3.6 using 25~Hz of line broadening and identical phasing parameters. 

\textsuperscript{89}Y measurements were done on a 3.2~mm HX MAS probe with 3.2~mm single cap sapphire rotors packed in the same way, spinning at 10~kHz under dry nitrogen. Due to the very long T\textsubscript{1} relaxation times of \textsuperscript{89}Y, a previous acquisition on undoped LYC had a very poor signal to noise ratio after 24~hours of measurements with a 300~s recycle delay. \ce{Sm^{3+}} doping drastically reduced the T\textsubscript{1}, allowing measurements to be acquired with a 120~s recycle delay. Spectra were referenced to 1M \ce{YCl3} at 0 ppm. Pulse lengths were calibrated on the same solution. Direct excitation was done with a single pulse experiment using a flip angle of 40$\degree$ corresponding to a 6~$\mu$s pulse at 100~W. All spectra were processed with Topspin 3.6 using 100~Hz of line broadening and identical phasing parameters.

PFG-NMR measurements were acquired at 7.05~T (300~MHz for \textsuperscript{1}H) on a super wide bore Bruker magnet equipped with an Avance III console and using a Diff50 probe with a 10~mm \textsuperscript{7}Li coil. Samples were loaded into 4~mm zirconia rotors and sealed under \ce{Ar} inside of a 5~mm valved NMR tube. A continuous 800~L hr\textsuperscript{–1} flow of \ce{N2} gas over the NMR tube maintained an inert atmosphere throughout each measurement and also regulated the sample temperature. The exact temperature of the probe was calibrated using a dry ethylene glycol solution in the same configuration. \textsuperscript{7}Li resonances were excited with 90$\degree$ pulses of length 15.5~$\micro$s at 100~W and referenced to 1M \ce{LiCl} at 0~ppm. T\textsubscript{1} relaxation times at each temperature were determined with a saturation recovery; recycle delays for PFG measurements were set to 2.5 T\textsubscript{1}. PFG-NMR experiments were conducted with a variable magnetic field gradient sequence [maximum gradient of 0.28~T~cm\textsuperscript{–1} (2800~G~cm\textsuperscript{–1})] and a diffusion sequence with a stimulated echo to protect against transverse (T\textsubscript{2}) relaxation. Gradient durations ($\delta$), diffusion times($\Delta$), and gradient strengths were selected to guarantee an adequate decay curve. All measurements used the greatest allowable gradient strength to minimize values of $\delta$ and $\Delta$ and maximize the signal to noise ratio. $\delta$ and $\Delta$ times never surpassed 10~ms and 100~ms, respectively. 

\subsection{Scanning Electron Microscopy}
High resolution images for ball milled and solid-state prepared LYC samples were obtained using a ThermoFisher Apreo C LoVac scanning electron microscope (SEM). Powdered samples were dispersed on top of carbon tape secured to a mounting stub. The samples sealed under Ar until they were transferred inside the SEM, where vacuum was pulled immediately to minimize exposure to the atmosphere. The images  were acquired under vacuum (below 10$^{-5}$~mbar) using an accelerating voltage of 5~keV with 0.40~nA of current at a working distance of 9.7~mm. Back-scattered and secondary electrons were detected with an Everhart-Thornley Detector (ETD). 

\subsection{Computational Methods}
\label{sec:compdetails}

We investigate the disorder of Li and Y in the \ce{Li3YCl6} structure using DFT\cite{kohnSelfConsistentEquationsIncluding1965}, where the  exchange and correlation term was approximated by the strongly constrained and appropriately normed (SCAN) functional\cite{sunStronglyConstrainedAppropriately2015}, as implemented in the Vienna \emph{ab initio} simulation package VASP code.\cite{kresseEfficiencyAbinitioTotal1996, kresseEfficiencyAbinitioTotal1996} In VASP, the wavefunctions are expanded in terms of  plane-waves with an energy cutoff of 520~eV, whereas the core electrons were described by the projector augmented-wave (PAW) theory.\cite{kresseUltrasoftPseudopotentialsProjector1999} The PAW potentials were: Li [$1s^{2}2s^{1}$], Y [$4s^{2}4p^{6}5s^{2}4d^{1}$], and Cl [$3s^{2}3p^{5}$]. Using a constant $k$-point density of 727 \AA$^{-1}$ across all calculations, the total energy was converged within 10$^{-5}$ eV, the atomic forces and stresses were converged to within 10$^{-2}$~eV~\AA$^{-1}$ and 0.29~GPa. These parameters were employed to investigate both the Li/Y orderings in the bulk structure, the stacking fault models, as well as the \ce{Li+} migration barriers. 

The low vacancy limit model was used to study \ce{Li+}  migration in \ce{Li3YCl6} structure, where the barriers for the migration of an isolated Li-vacancy are evaluated based on the nudged elastic band (NEB) method\cite{henkelmanClimbingImageNudged2000,sheppardOptimizationMethodsFinding2008} available in VASP. NEB forces are converged within 0.50~eV~\AA$^{-1}$, following the protocol developed by Chen \emph{et al.}\cite{chenIonicTransportPotential2019} The supercell models of \ce{Li3YCl6}  used in NEB calculations introduced a minimum distance of $\approx$10~\AA{} between the migrating \ce{Li+} ions, which avoids any undesired image-image interaction. The introduction of Li vacancies to enable ion migration was compensated by a counter charge in the form of a jellium.

NMR parameters for $^{6,7}$Li and $^{89}$Y were calculated using the CASTEP.\cite{clarkFirstPrinciplesMethods2005} The generalized gradient approximation by Perdew, Burke, and Ernzerhof \cite{perdewGeneralizedGradientApproximation1996} was employed to approximate the exchange-correlation term. All calculations were run with ''on-the-fly`` ultrasoft pseudopotentials as supplied by CASTEP. NMR calculations were run using the projector augmented-wave method (GIPAW).\cite{pickardAllelectronMagneticResponse2001, yatesCalculationNMRChemical2007}  The scalar-relativistic zeroth-order regular approximation (ZORA)\cite{yatesRelativisticNuclearMagnetic2003} was used for relativistic effects. The procedure for each calculation followed an identical sequence of single-point energy convergence with respect to a plane-wave cutoff energy and $k$-point grid, followed by a geometry optimization using the 
Broyden–Fletcher–Goldfarb–Shanno (BFGS) algorithm without constraints on the lattice, atomic positions, or imposed symmetry, and concluded with a convergence of the chemical shielding constants with respect to a new cutoff energy and $k$-point grid. 

Convergence criteria were the following: for single-point energy calculations, energy convergence tolerance was 0.5 meV~atom$^{-1}$; for geometry optimization, energy convergence tolerance was 0.02~meV~atom$^{-1}$, maximum ionic force tolerance was 0.05~eV~\AA$^{-1}$, maximum ionic displacement tolerance was 0.001~\AA, and maximum stress component tolerance was 0.1~GPa. For the NMR calculations, the isotropic chemical shielding constant convergence tolerance was 0.01~ppm for Li shifts and 0.5~ppm for Y shifts. Converged cutoff energies and $k$-point grids used for geometry optimization and NMR parameter calculations are tabulated in Section~S4 of the SI.

CASTEP-calculated isotropic chemical shielding constants were converted to experimentally-relevant chemical shifts through a semi-empirical linear regression according to Eq.~\ref{eq:NMRshift}
\begin{equation}
\label{eq:NMRshift}
\delta_{iso}=\sigma_{ref}-m \times \sigma_{iso}
\end{equation} 
where $\delta_{iso}$ is the experimentally calibrated chemical shift, $\sigma_{iso}$ is the computed isotropic chemical shielding constant, $m$ is a scaling factor, and $\sigma_{ref}$ is the computed isotropic chemical shielding constant of a reference compound. Calibration curves were constructed using a series of binary \ce{Li} and \ce{Y}-containing compounds, in a method analogous to that reported by Sadoc \emph{et al.} for \textsuperscript{19}F.\cite{sadocNMRParametersAlkali2011} Final calibration curves, computed binaries, and references for $^{6,7}$Li and $^{89}$Y shifts are reported in the supporting information (see Sections~S5, S10).

\subsection{Electrochemical Impedance Spectroscopy}
Ionic conductivity measurements were carried out with AC impedance on a Solartron 1260A analyzer. Pellets of $\approx$90 mg of powder were uniaxially pressed for 15 min at 350 MPa to relative densities of 76-79\% and loaded into a custom-design hermetically sealed cell with stainless steel electrodes and a PEEK housing. The cell was placed in a Tenney TJR environment chamber and maintained under 180 MPa of pressure throughout the measurements with a vice. Impedance spectra were collected with excitation voltages of 10~mV - 30~mV between 8 MHz and 100 Hz and at a series of temperatures from 263 K to 313 K in 10 K increments. Fitting of EIS spectra was conducted with the ZView software package. Errors in the calculated conductivity and activation energy accounted for fitting, temperature, and thickness measurement errors.

\begin{acknowledgement}
This work made use of the shared facilities of the UC Santa Barbara MRSEC (DMR 1720256), a member of the Materials Research Facilities Network (http://www.mrfn.org) and the computational facilities administered by the Center for Scientific Computing at the CNSI and MRL (an NSF MRSEC; CNS 1725797, DMR 1720256). Use was made of computational facilities purchased with funds from the National Science Foundation (CNS-1725797) and administered by the Center for Scientific Computing (CSC). The CSC is supported by the California NanoSystems Institute and the Materials Research Science and Engineering Center (MRSEC; NSF DMR 1720256) at UC Santa Barbara.  Powder X-ray diffraction data were collected on beamline 17-BM at the Advanced Photon Source at Argonne National Laboratory, which is supported by the U.S.\ Department of Energy, Office of Science, Office of Basic Energy Sciences under Contract No. DEAC02-06CH11357. We gratefully acknowledge Dr. Benjamin Trump for data collection. Neutron diffraction data was collected at the National Institute of Standards and Technology Center for Neutron Research (NCNR). This material is based upon work supported by the National Science Foundation Graduate Research Fellowship under Grant No. 1650114. H.A.E.\ thanks the National Research Council (USA) for financial support through the Research Associate Program. This work was partially supported by the RISE internship program through the MRSEC Program of the National Science Foundation under Award No.\ DMR 1720256. P.C.\ acknowledges funding from the National Research Foundation under his NRF Fellowship NRFF12-2020-0012, as well as support from the Singapore Ministry of Education Academic Fund Tier 1 (R-284-000-186-133). The computational work for this article was partially performed on resources of the National Supercomputing Centre, Singapore (https://www.nscc.sg). Certain commercial equipment, instruments, or materials are identified in this document. Such identification does not imply recommendation or endorsement by the National Institute of Standards and Technology, nor does it imply that the products identified are necessarily the best available for the purpose. This work was performed, in part, at the William R. Wiley Environmental Molecular Sciences Laboratory, a national scientific user facility sponsored by US Department of Energy, Office of Biological and Environmental Research and located at PNNL. PNNL is operated by Battelle for the US Department of Energy under contract DE-AC05-76RLO1830.

\end{acknowledgement}

\begin{suppinfo}
Cryo-TEM images; Stacking fault model construction, neutron diffraction patterns, high temperature SS-LYC X-ray diffraction patterns, fit for $\beta$-LYC; accuracy of exchange and correlation functional; converged $k$-point and planewave energy cutoffs for NMR calculations; $^6$Li and $^{89}$Y NMR calibration curves; $^6$Li, $^{89}$Y NMR and XRD on undoped and Sm-doped LYC; fit for SS-LYC $^{89}$Y NMR spectrum; EIS spectra and fits; \ce{Li+} self-diffusion coefficients as measured by PFG-NMR; SEM images; variable diffusion time PFG-NMR; bond valence maps on pristine and faulted structures.
\end{suppinfo}

\bibliography{References}

\clearpage

\begin{figure*}[ht!]
\includegraphics[width=6in]{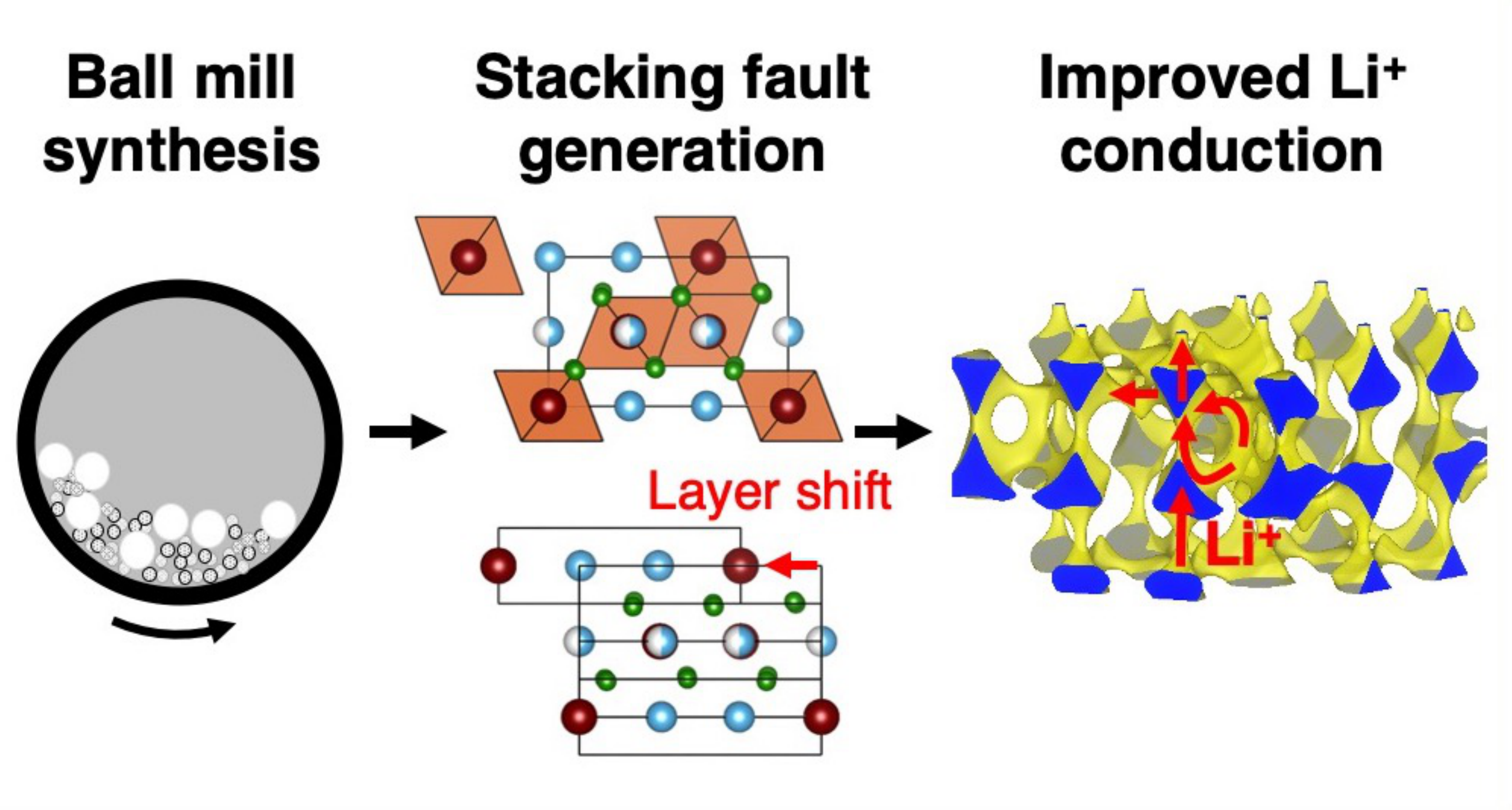}
  \caption{
  For Table of Contents Only 
  \label{fig:TOC_fig}
  }
\end{figure*}

\end{document}